\documentclass[%
reprint,
nofootinbib,
amsmath,amssymb,
aps,
prl
]{revtex4-2}

\usepackage{soul,xcolor}
 \usepackage{mathtools}
\usepackage{amsmath}

\usepackage{braket}
\usepackage{lineno}
\renewcommand{\vec}[1]{\boldsymbol{#1}}

\begin{document}

\title{Nanoimprinted Exciton-Polaritons Metasurfaces: Cost-Effective, Large-Scale, High Homogeneity, and Room Temperature Operation}

%%% ================================================================================
	%%% AUTHOR AND AFFILIATIONS 
	%%% ================================================================================ 
	\author{Nguyen Ha My Dang$^{1}$} 
	\author{Paul Bouteyre$^{2}$} 
	\author{Ga\"{e}lle Tripp\'e-Allard$^3$}
	\author{C\'eline Chevalier$^{1}$}  
	\author{Emmanuelle Deleporte$^3$}
	\author{Emmanuel Drouard$^1$}
	\author{Christian Seassal$^1$}  
	\author{Hai Son Nguyen$^{1,4}$}     
	\email{hai-son.nguyen@ec-lyon.fr} 
	\affiliation{$^1$ Univ Lyon, ECL, INSA Lyon, CNRS, UCBL, CPE Lyon, INL UMR 5270, 69130 \'{E}cully, France}  
	\affiliation{$^2$ Department of Physics and Astronomy, University of Sheffield, S3 7RH, Sheffield, UK}
	\affiliation{$^3$ Lumi\`ere, Mati\`ere et Interfaces (LuMIn) Laboratory, Universit\'e Paris-Saclay, ENS Paris-Saclay, CNRS, CentraleSup\'elec, 91190 Gif-sur-Yvette, France}
	\affiliation{$^4$ Institut Universitaire de France (IUF), 75231 Paris, France} 
	
\begin{abstract}
Exciton-polaritons represent a promising platform that combines the strengths of both photonic and electronic systems for future optoelectronic devices. However, their application is currently limited to laboratory research due to the high cost and complexity of fabrication methods, which are not compatible with the mature CMOS technology developed for microelectronics. In this work, we develop an innovative, low-cost, and CMOS-compatible method for fabricating large surface polaritonic devices. This is achieved by direct patterning of a halide-perovskite thin film via thermal nanoimprint. As a result, we observe highly homogeneous polaritonic modes of quality factor $Q\approx 300$ at room temperature across a centimetric scale. Impressively, the process provides high reproducibility and fidelity, as the same mold can be reused more than 10 times to imprint the perovskite layer on different types of substrates. Our results could pave the way for the production of low-cost integrated polaritonic devices operating at room temperature.
\end{abstract}

	\maketitle
 
%%%%%%%%%%%%%%%%%%%%%%%%%%  body  %%%%%%%%%%%%%%%%%%%%%%%%%%
\section{Introduction}
Over the past decade, exciton-polaritons have emerged as a promising platform for the next generation of all-optical devices. These hybrid half-matter, half-light quasi-particles arise from the strong coupling regime between excitonic excitations in semiconductors, or excitons, and confined photons in optical cavities\cite{Claude1994}. Owing to their hybrid nature, these quasi-particles exhibit low losses and high-speed propagation across long distances via the photonic component, while also demonstrating strong nonlinearity inherited from the excitonic component\cite{QuantumFLuidsOfLight}. Although various polaritonic devices, such as polaritonic memories, transistors, gates, and diodes, have been experimentally demonstrated by different groups, these devices remain in the realm of laboratory research\cite{Sanvitto2016}. This fascinating platform faces two significant challenges: much of its development has been done with GaAs semiconductors requiring operation at cryogenic temperatures for excitons, and the device fabrication demands costly processes that are incompatible with CMOS technology.

In an effort to make polaritonic devices suitable for practical applications, research has sought out room-temperature excitons and thus, materials with high excitonic binding energy such as GaN,\cite{Christopoulos2007,Daskalakis2013} ZnO,\cite{Franke_2012} organic semiconductors,\cite{Lidzey1998,Plumhof2014} and more recently, monolayers of transition metal dichalcogenides,\cite{Liu2014,Grosso2017} and halide perovskites.\cite{Su2021} On the other hand, to implement device functionality, the photonic confinement of polaritons often needs to be harnessed via micro/nano-structuration. For instance, directly patterning 2D hybrid organic-inorganic perovskite into sub-wavelength scale lattices has given rise to a novel polaritonic system, which we refer to as a polariton metasurface, wherein the excitonic material itself hosts the photonic confinement.\cite{Dang2020,Lu2020,Dang2022,Kim2021} The metasurface approach has allowed for the exploration of uncharted polaritonic regimes generated based on the collective behaviors of sub-wavelength active elements. Within this framework, dispersion tailoring of energy–momentum band structure with linear, parabolic, and multi-valley dispersion characteristics has been demonstrated at room temperature.\cite{Dang2020} Furthermore, it has been shown that the concepts of photonic bound states in the continuum can be transferred to polaritonic states in a polariton metasurface to enhance the polariton quality factor and engineer polarization singularities.\cite{Dang2022,Kim2021}  It is worth noting that since the initial proposals for two-dimensional (2D) hybrid perovskites, the polariton metasurface approach has been subsequently applied to the GaAs platform at cryogenic temperatures,\cite{Ardizzone_Nature2022,gianfrate2023optically,riminucci2023boseeinstein} and then to organic semiconductors, \cite{Berghuis2023}, monolayers of transition metal dichalcogenides,\cite{Kravtsov2020,Maggiolini2023} and three-dimensional (3D) all-inorganic perovskite materials at room temperature,\cite{wu2023roomtemperature}  resulting in the Bose-Einstein condensation of polaritons from bound states in the continuum. However, the fabrication of such structures still requires high-cost processes (such as epitaxial growth, electronic beam lithography, and plasma-assisted etching) or low-reproducibility ones (like exfoliation and infiltration of excitonic materials into pre-patterned structures).

In this study, we present a cost-effective and CMOS-compatible technique to create large-scale polaritonic devices. We accomplish this by directly patterning 2D hybrid organic-inorganic perovskite thin films via thermal nanoimprint lithography. Notably, the same mold can be effectively used to imprint a variety of 2D perovskites on different substrates, providing versatile application potential. The imprinting process triggers recrystallization of the perovskite layer under pressure, enhancing the structural integrity and performance of the device. The result is remarkably homogenous polaritonic modes of a mere few meV linewidth, observable at room temperature across centimetric expanses. A notable feature of our method is its long-term stability; the emission of polaritonic modes remains robust and undiminished even after 8 months. This approach could potentially herald the advent of cost-effective, integrated polaritonic devices functioning at room temperature, with promising implications for optoelectronic applications.

\section{Fabrication methods}
\begin{figure}[htbp]
	\begin{center}
		\includegraphics[width=0.5\textwidth]{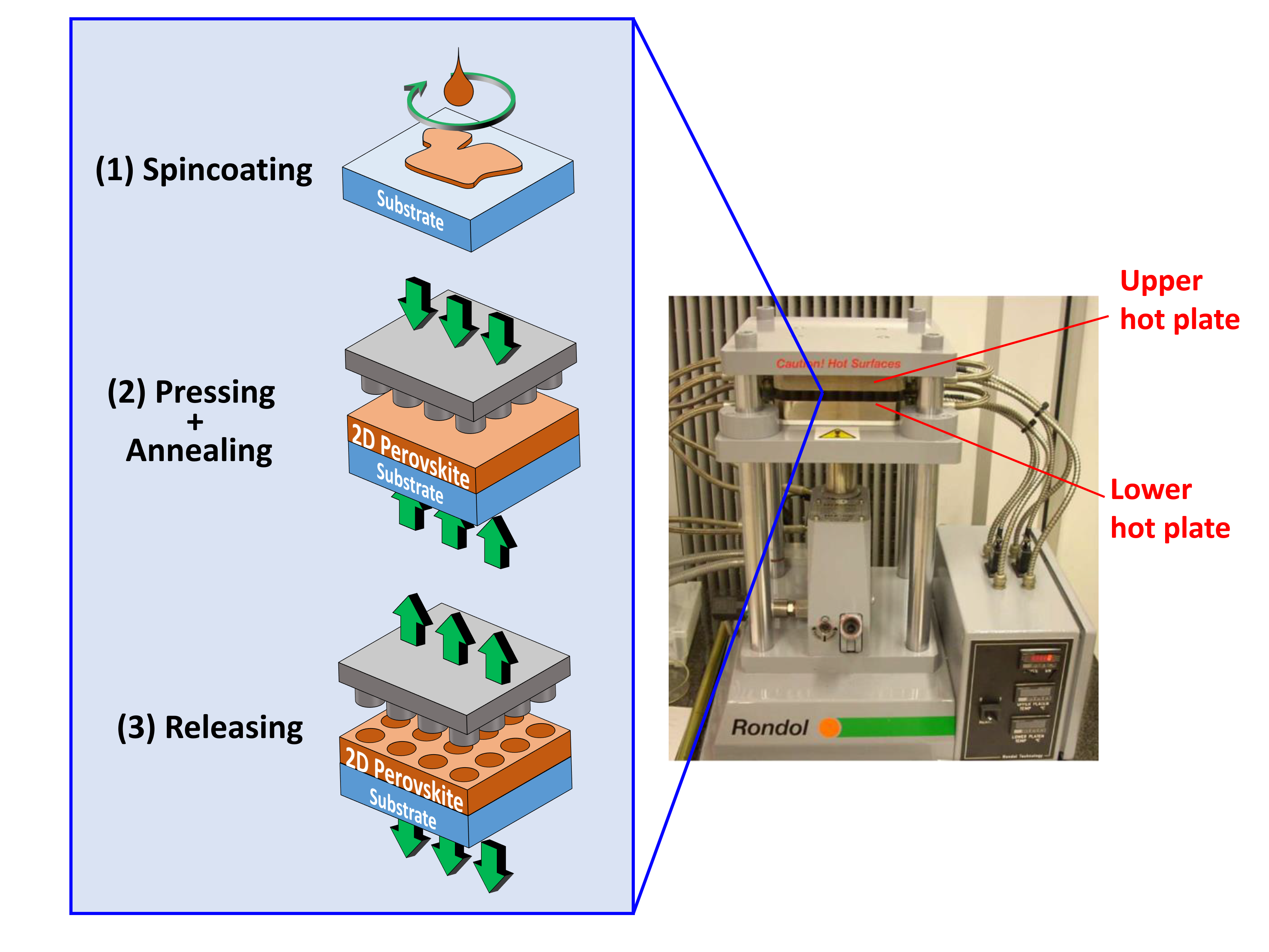}
	\end{center}
	\caption{Schematic of the fabrication steps to make a polariton metasurface via thermal nanoimprint. Step (1), spin-coating for thin-film perovskite deposition, is performed in a glove box filled with nitrogen. Steps (2) and (3), molding and recrystallizing the perovskite layer, are performed in an ambient atmosphere with the help of a hydraulic press (photo on the right).\label{fig1}}
\end{figure}

Figure \ref{fig1} depicts our fabrication method for creating a polariton metasurface. Briefly, a thin film of 2D hybrid perovskite is deposited on a substrate via spin-coating. Next, a mold with periodic nanostructures is placed on top of the deposited perovskite substrate. Then, high pressure and temperature are applied, transferring the pattern to the perovskite film. We note that this fabrication method is adapted from a similar one that is developed to pattern 3D hybrid perovskite.\cite{Mermet2023} The different steps of the fabrication process are detailed below.

\subsection{Perovskite materials and deposition}
Our excitonic material is 2D hybrid organic-inorganic perovskites, one of the most prominent materials for studying exciton-polaritons at room temperature, thanks to remarkable excitonic features despite an extremely simple fabrication method. Indeed, high-crystal-quality perovskite thin films made of alternating organic/inorganic monolayers can be obtained by a single-step spin-coating process, followed by thermal annealing. These alternating monolayers correspond to a multi-quantum well structure, with the organic layers playing the role of potential barriers (approximately 1 nm) that confine electrons and holes within approximately 0.5 nm of inorganic quantum wells. Moreover, the confinement effect is strengthened thanks to the high dielectric contrast between the organic and inorganic layers. The combination of quantum confinement and dielectric confinement results in high oscillator strength as well as high exciton binding energy, ranging up to hundreds of meV at room temperature.

Here we developed the thermal nanoimprint of two different types of 2D hybrid perovskite: bi-(butylammonium) tetraiodoplumbate, known as BAPI, with the chemical formula (C$_4$H$_9$NH$_3$)$_2$PbI$_4$; and bi-(phenethylammonium) tetraiodoplumbate, known as PEPI, with the chemical formula (C$_6$H$_5$C$_2$H$_4$NH$_3$)$_2$PbI$_4$. At first, a thin layer of perovskite (PEPI or BAPI wt 20$\%$ DMF solution) is deposited on the substrate by spin-coating at 5000 rpm for 30 sec. Then the film is annealed at 95°C for 90s for crystallization before undergoing recrystallization during the imprinting process. This deposition takes place in a glove box filled with nitrogen. We note that the surface treatment of the substrate plays a vital role in perovskite deposition. Here, the substrates are dipped in solvents (acetone, ethanol, isopropanol – IPA) in an ultrasonic bath for 15 minutes for each step and then processed in a UV- ozone box for 20 minutes to enhance the surface energy and improve the homogeneity of perovskite deposition.

\subsection{Mold fabrication and preparation}
We used laser interference lithography (LIL) to define the designed patterns on the silicon molds (1$\times$1 to 2$\times$2 cm$^2$ size). This low-cost lithography technique allows for obtaining large surfaces of homogeneous periodic patterns with high resolution. See \cite{Liu2015} for further details on our optimized LIL process. After lithography, the patterns were transferred by reactive ion etching (RIE) into the silicon substrate. Before imprinting, the molds are cleaned by acetone, ethanol, IPA, and then their surface is functionalized by a silanization process to avoid sticking with perovskite. A solution is prepared by dissolving a few mg of FDDTS (Perfluorododecyltrichlorosilanes – Sigma Aldrich) into 5 ml of n-heptane. The molds are then left inside this solution for silanization for 30 minutes. During the silanization treatment, the –OH groups bonded to Si particles on the surface of the molds interact with the –SiCl$_3$ groups inside the FDDTS and form a bond Si (surface) – O – Si (FDDTS). The mold surface is therefore terminated by silane and its functional organic group, which reduces the surface energy or increases the surface tension. Finally, the molds are cleaned with acetone and baked at 100°C for 20 minutes.
\begin{figure*}[htbp]
	\centering\includegraphics[width=\textwidth]{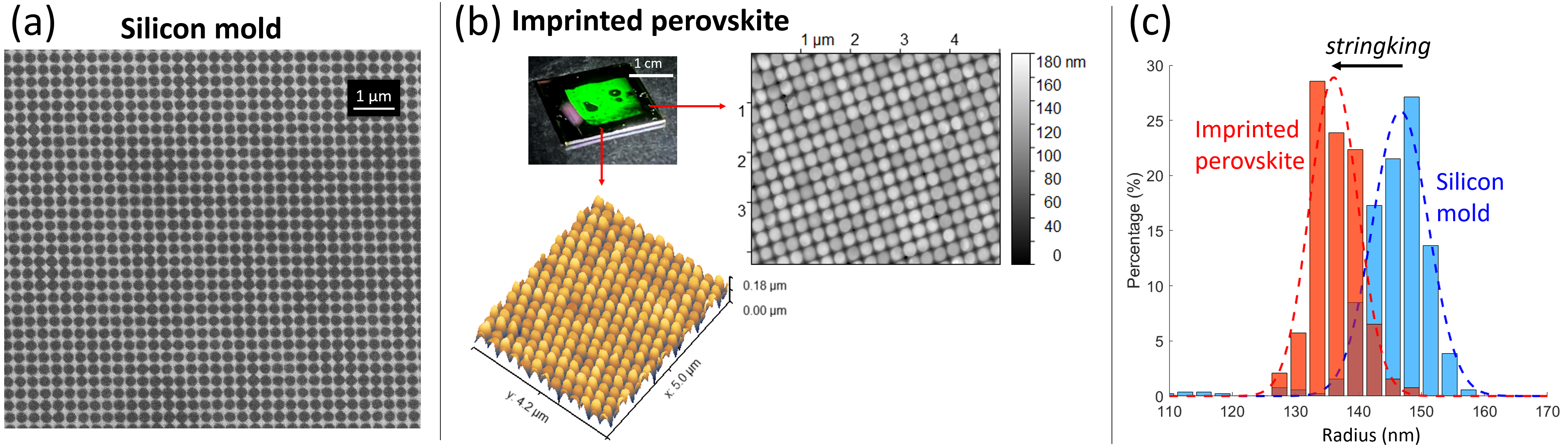}
	\caption{(a) SEM image of the silicon mold. (b) Photo (top left) and AFM images of the imprinted BAPI layer on silicon substrate. (c) Histogram of the hole (pillar) radius in the silicon mold (imprinted BAPI).\label{fig2}} 
\end{figure*}
\subsection{Perovskite deposition and thermal imprinting}
A manual table-top hydraulic press (see Fig.~\ref{fig1}), containing two stainless steel plates integrated with a heating system (up to 300°C for each plate), is employed for the imprint. This press's upper plate is fixed and the force is applied by manually lifting the lower plate toward the top one. The imprinting process is operated in an ambient atmosphere. The deposited perovskite thin film is settled between the two plates inside the thermal press. After that, a silicon mold with the desired patterns is positioned on top of the perovskite film. The lower plate is lifted so that the force is gradually increased until the pressure reaches the value of around 100 bar. Once reaching the pressure, the temperature of the two metal plates in contact with the sample and silicon mold is set to 100°C. This process takes around 10 minutes for the temperature to stabilize. The sample and the mold are continually pressed (constant force) and heated (at 100°C) for 15 minutes. This process would recrystallize the perovskite layer and at the same time mold this layer into the subwavelength metasurface. After 15 mins, the heat is turned off. The pressure is released once the temperature is cooled down to 35°C. Finally, the imprinted sample is encapsulated with 200\,nm-thick PMMA layer to avoid contact with humidity and moisture.

\section{Results and discussion}
\subsection{Pattern transfer from silicon mold to thin film perovskite}\label{sec:pattern_fidelity}
Figure~\ref{fig2}a presents the Scanning Electron Microscopy (SEM) image of the 1.85$\times$1.85 cm$^2$-size silicon mold having square lattice of holes of 305 nm period and 200 nm depth. We first use this mold to pattern a BAPI layer on a silicon substrate. The diffraction due to the periodic pattern on the imprinted perovskite is nicely observed (see the photo in top of Fig.~\ref{fig2}b), confirming the success pattern transfer over a centimetric scale. Two defects regions, visible on the photos, are due to some dusts that get onto the perovskite surface just before the imprinting. The SEMs images of the patterned BAPI surface (see Fig.~\ref{fig2}b) provide different zooms within a 300$\times$200 µm$^2$ area of the imprinted sample. From the SEM characterization, we confirm that the period of the BAPI pillar lattice is consistent with the period of the original mold. However, the SEM images also indicate that pattern size of BAPI pillars is smaller than that of the mold air holes. For a quantitative analysis, we extracted the size distribution histogram of holes and pillars from SEM images (see Fig.~\ref{fig2}c). These results show two gaussian statistics, with PEPI pillars radius is 136 nm $\pm$ 5 nm and holes radius is 146 nm $\pm$ 5 nm. The mean value of the imprinted pillar radius is therefore 10 nm smaller than the hole radius of the template mold. The size difference confirms that BAPI recrystallized and shrank after the imprinting process. We note that despite a shift of average values, the two distributions have the same standard deviation. This implies that the mold pattern has been transferred to the perovskite pattern with high fidelity.
\subsection{Fidelity of polaritonic modes in imprinted perovskite layers on different substrates}
\begin{figure*}[htbp]
	\centering\includegraphics[width=\textwidth]{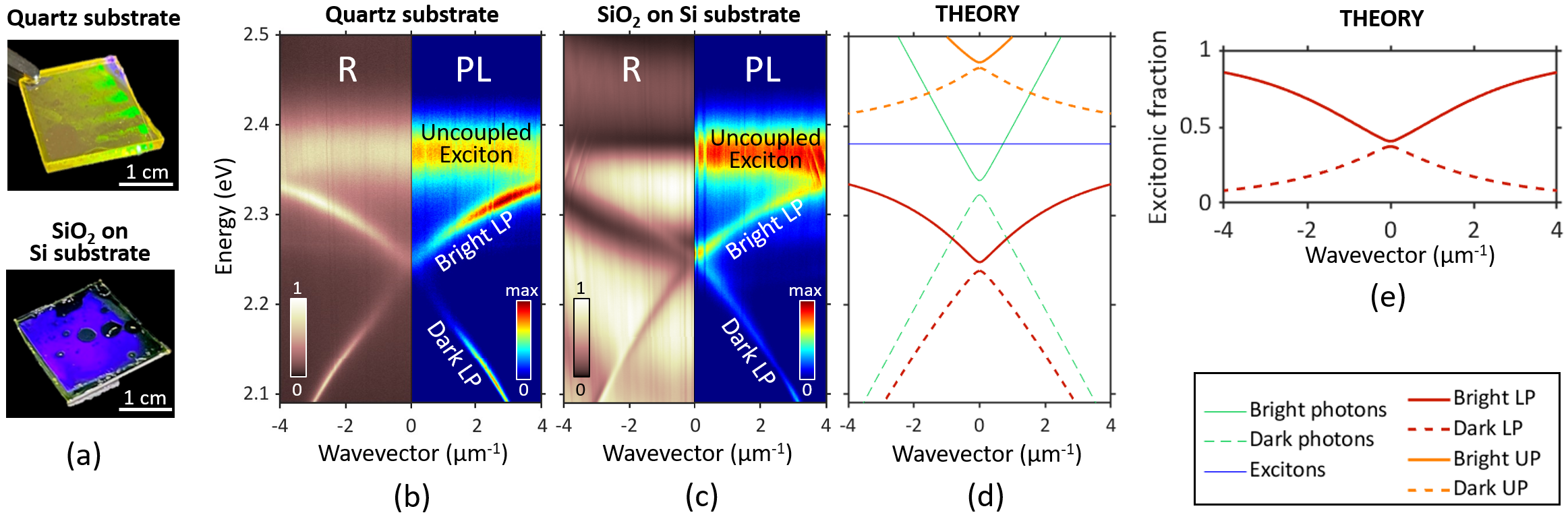}
	\caption{(a)  Photo of the imprinted BAPI layer on two different substrates. (b,c) Experimental results of the angle-resolved reflectivity (left panels) and angle-resolved photoluminescence (right panels) for the imprinted BAPI layer on (b)  quartz substrate and on (c) 2 µm of SiO$_2$ on silicon substrate. The experiments are performed in s-polarization. (d) Theoretical calculations of photonic Bloch resonances and polaritonic modes. (e) Theoretical calculations of the excitonic fractions corresponding to the Bright LP and Dark LP branches. \label{fig3}} 
\end{figure*}

We now turn our attention to the formation and fidelity of polaritonic modes in perovskite metasurfaces, which are imprinted from the same mold onto different substrates. The mold depicted in Fig. \ref{fig2}a of the previous section is used for patterning two layers of BAPI. The first layer is pre-deposited on a quartz substrate, and the second is on a substrate comprising 2 µm of SiO$_2$ on a silicon base (see Fig. \ref{fig3}a). To demonstrate the strong coupling regime and characterize the polaritonic modes in the fabricated devices, we perform angle-resolved micro-reflectivity and micro-photoluminescence. The experimental setup is detailed in the Appendix, section \ref{sec:Setup}. It is important to note that in angle-resolved spectroscopy along the $\Gamma X$ direction of the Brillouin zone, the strong coupling regime is revealed only in s-polarization measurements, since the polaritonic modes in p-polarization are non-dispersive (further details are provided in the Appendix, section \ref{sec:modelVSexp}). Consequently, unless otherwise specified, only s-polarization results are presented hereafter.

 \begin{figure*}[htbp]
 	\centering\includegraphics[width=0.8\textwidth]{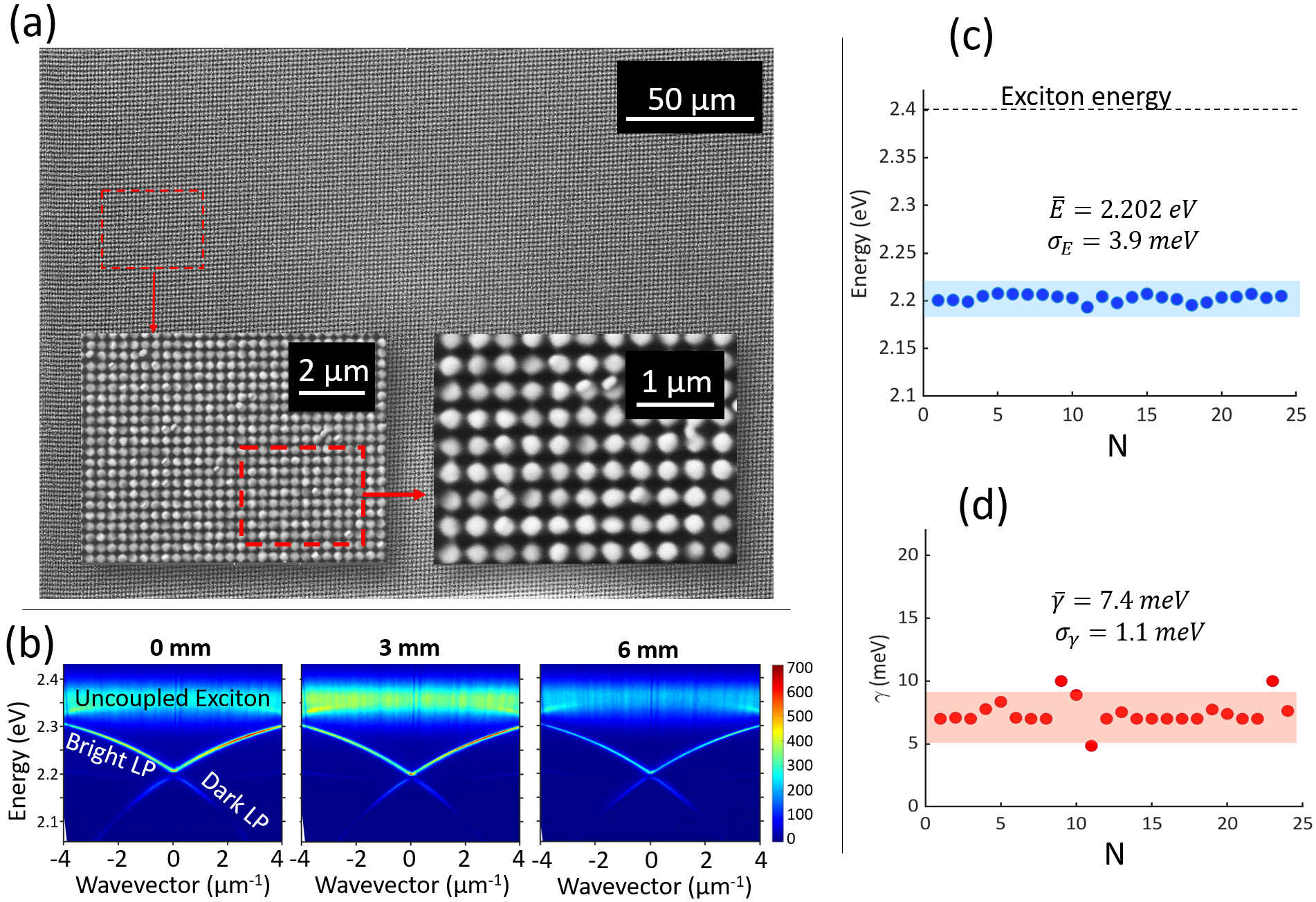}
 	\caption{(a) SEM image of imprinted PEPI metasurface. (b) Angular resolved PL measured in three different positions on imprinted PEPI metasurface. (c,d) Energy peaks and linewidth of polaritonic modes extracted from ARPL measured at different positions on patterned PEPI metasurface.\label{fig4}} 
 \end{figure*}

Figures \ref{fig3}b and c display the experimental outcomes from the samples with a quartz substrate (Fig.\ref{fig3}b) and a SiO$_2$ optical spacer on a silicon substrate (Fig.\ref{fig3}c), respectively. In both sets of results, an angle-independent signal at 2.38 eV, corresponding to uncoupled excitons, is observed. More importantly, two dispersive polaritonic modes are noted when examining the resonances in the measured spectra: a bright and broad lower-polaritonic mode, termed "Bright LP," and a sharp, dim lower-polaritonic mode, termed "Dark LP." These bands stem from the strong coupling between photonic Bloch resonances arising from first-order diffraction along $\Gamma X$ and perovskite excitons. The development of these modes is elaborated in the Appendix, section \ref{sec:model}). In simpler terms, they resemble the bright and dark polaritonic branches in 1D polaritonic gratings with perovskite- \cite{Lu2020} and GaAs \cite{Ardizzone_Nature2022,gianfrate2023optically, riminucci2023boseeinstein}-based metasurfaces. The strong coupling regime is unequivocally demonstrated in both reflectivity and photoluminescence results by the bending of the Bright LP branch as it approaches the exciton energy, indicative of the anticrossing effect. This is observed by monitoring the polaritonic dispersion as a function of the in-plane wave vector. Like other room-temperature polaritonic systems with high bandgap materials, the signal of upper polariton branches is not detected due to BAPI's strong absorption above the excitonic transition. A minor but noteworthy detail is the modulation of the background signal in the reflectivity results of Fig.~\ref{fig3}c, attributable to the presence of Fabry-Perot modes within the 2 µm SiO$_2$ optical spacer.
 
 Remarkably, the dispersion of polaritonic modes from both reflectivity and photoluminescence signals is nearly identical for the two samples. This demonstrates the high fidelity of the polaritonic modes achieved using the same mold and perovskite but on different substrates. Furthermore, the polariton branches observed in both samples align well with theoretical predictions (see Fig.\ref{fig3}d). By comparing the experimental results with theoretical models, we deduce a theoretical Rabi splitting energy of 210 meV, aligning well with previous reports on perovskite metasurfaces fabricated using the infiltration method \cite{Dang2020,Dang2022}. Finally, Figure\ref{fig3}e presents the excitonic fraction of Bright LP and Dark LP, illustrating that the observed polaritonic modes are genuine light-matter mixed states. The ratio of exciton fraction/ photonic fraction ranges from 8$\%$/92$\%$ to 85$\%$/15$\%$ across our wave-vector scanning range. At the $\Gamma$ point (i.e. zero-wavevector), the Bright LP and Dark LP exhibits exciton fraction/ photonic fraction ratio of 41$\%$/59$\%$ and 38$\%$/62$\%$ respectively.

\subsection{Homogeneity of polaritonic modes over centrimetric scale}
To investigate the homogeneity of polaritonic modes across the imprinted samples, we perform angle-resolved micro-photoluminescence measurements over the entire sample area. This study focuses on PEPI metasurfaces rather than BAPI ones, as the stability of our PEPI layers is superior to that of BAPI layers, making them more suitable for systematic characterizations (the reproducibility and stability of PEPI metasurfaces will be discussed in the subsequent section \ref{sec:stability}). Figure~\ref{fig4}a presents SEM images at different magnifications of our PEPI metasurface, confirming the morphological homogeneity over a surface area of 300$\times$300 µm$^2$. Concerning the homogeneity of polaritonic modes, Fig.\ref{fig4}b displays three photoluminescence measurements taken at distances of 0 mm, 3 mm, and 6 mm from the sample's center. These results demonstrate almost identical polaritonic modes for both Bright LP and Dark LP. Quantitatively, measurements at 25 different positions, corresponding to 25 meshed coordinates within a 1$\times$1 mm$^2$ area, are conducted. From these measurements, we extract the polariton energy and the corresponding linewidth of the Bright LP at the $\Gamma$ point (i.e. zero wave vector), which are presented in Figs.\ref{fig4}c and d. These findings reveal a polariton energy of 2.202 eV with a fluctuation of 3.9 meV, and a linewidth of 7.4 meV with a fluctuation of 1.1 meV. This indicates homogenous polaritonic modes (as the energy fluctuation is smaller than the linewidth), with a quality factor $Q\approx 300$.

\subsection{Reproducibility and Stability}\label{sec:stability}
\begin{figure}[htbp]
	\centering\includegraphics[width=0.5\textwidth]{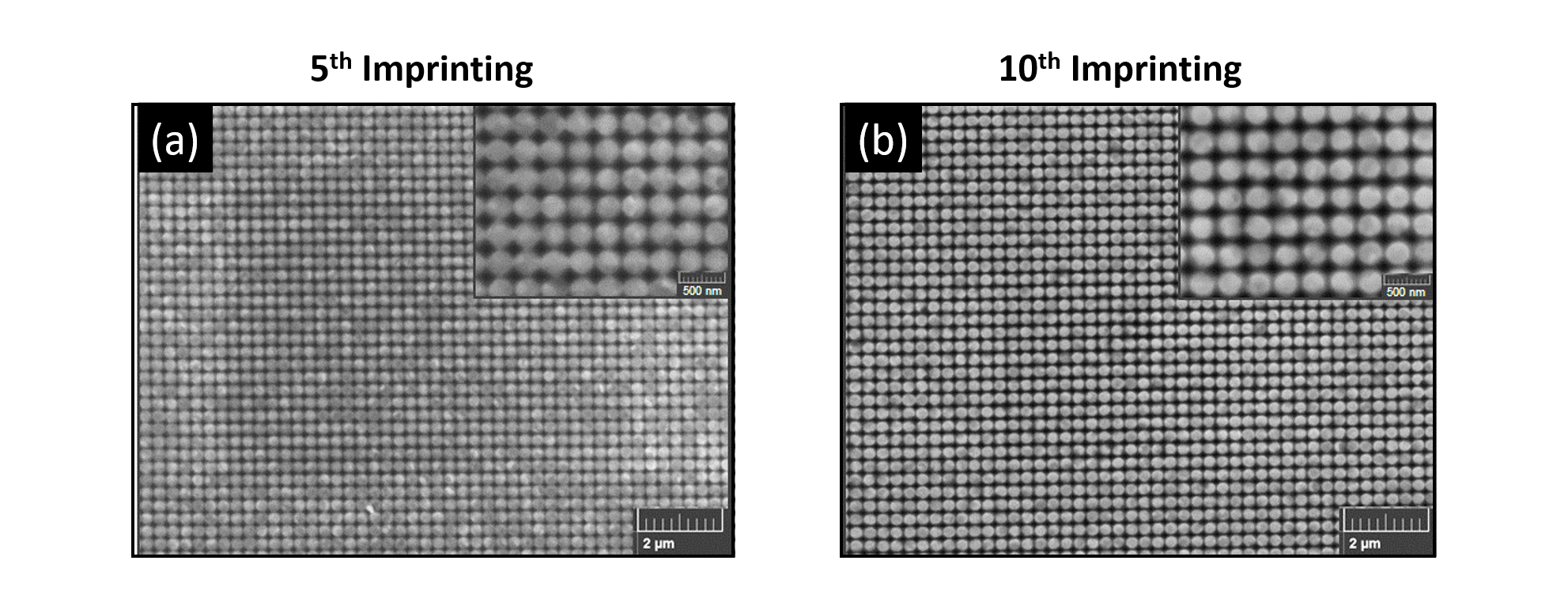}
	\caption{(a,b). SEM images of different imprinted PEPI metasurfaces reusing the same mold: (a) the 5\textsuperscript{th} imprinting sample and (b) the 10\textsuperscript{th} imprinting sample. \label{fig5}} 
	%[Note: fig 3, peak B: 2.157 eV; peak A: 2.279 eV; peak C: 2.787 eV ]
\end{figure}
Nanoimprinting is a highly efficient fabrication technique. A single nanopatterned mold can be used to produce multiple metasurfaces. By optimizing the imprinting parameters, one mold can be utilized more than ten times without any degradation in quality. As a demonstration, Figures~\ref{fig5}a and b compare PEPI metasurfaces created by imprinting the same mold under similar conditions. The two imprinted patterns are identical, even after 10 uses of the mold, thereby confirming the high reproducibility of the fabrication method.  

\begin{figure}[htbp]
	\centering\includegraphics[width=0.5\textwidth]{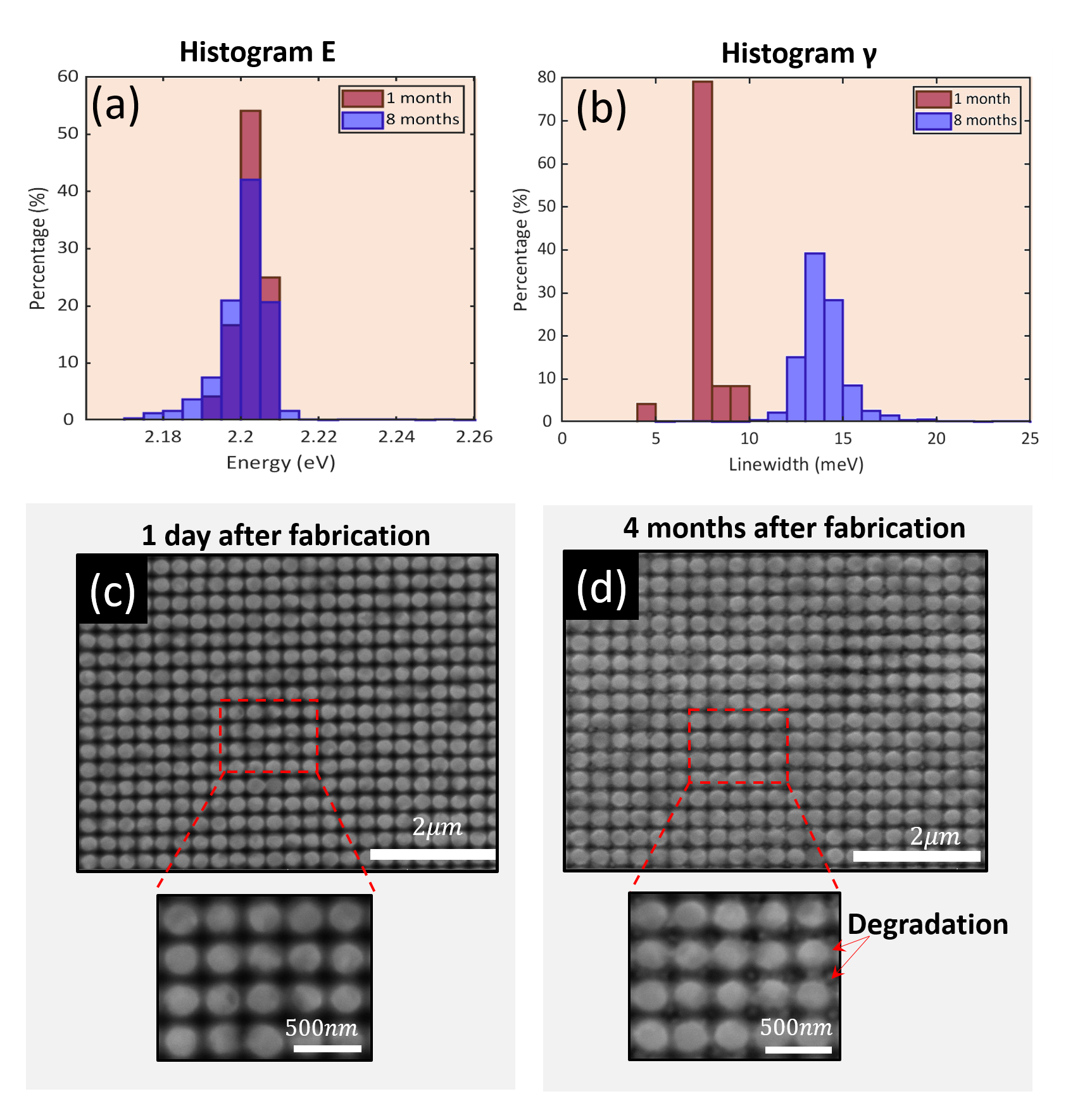}
	\caption{(a,b). SEM images imprinted PEPI metasurfaces 1 day and 4 months after fabrication. The sample was not covered by PMMA and stored inside a glove box.  (c,d) Histogram of polaritonic energy peaks (c) and linewidth (d) of PEPI metasureface after one month and eight months of fabrication. \label{fig6}} 
	%[Note: fig 3, peak B: 2.157 eV; peak A: 2.279 eV; peak C: 2.787 eV ]
\end{figure}

Under appropriate storage conditions, such as PMMA encapsulation and storage in a nitrogen-filled glove box, the lifetime of imprinted 2D perovskite metasurfaces can be significantly enhanced. To investigate the degradation of polaritonic modes, Figs\ref{fig6}a and b present histogram charts of the polaritonic energy peak and the polaritonic linewidth of the Bright LP at the $\Gamma$ point. These were derived from angle-resolved photoluminescence measurements conducted at various positions on the same PEPI metasurface (up to 1 cm apart, as mentioned in the previous section), taken one month and eight months after fabrication. In Figure ~\ref{fig6}a, the energy peaks of the polaritonic resonance barely shifted, demonstrating the stability of the metasurface (after 1 month: average energy $\overline{E}=2.202$ meV with a standard deviation $\sigma_E=3.9$ meV; after 8 months: $\overline{E}=2.201$ meV with $\sigma_E=6.7$ meV). Conversely, Figure ~\ref{fig6}b shows that the polaritonic linewidths broaden over time. The average polaritonic linewidth increased from $\overline{\gamma}=7.4\pm1.1 meV$ (measured after 1 month) to $\overline{\gamma}=14.1\pm3.3 meV$ (measured after 8 months), indicating a reduction in the quality factor from 300 to 150. These findings imply that although the shape of the polaritonic modes remains stable over time, the polariton lifetime gradually degrades. Future studies are required to develop strategies to preserve the quality factor of perovskite metasurfaces for long-term device applications.

Beside that, lifetime of perovskite metasurface reduced significantly without PMMA coating. SEM images in figure 6 show morphology of PEPI metasurface 1 day after fabrication and after 4 months storing inside glovebox. Just after 4 months, we started to see the signs of film degradation. There are several defects surrounding the patterns which could be the formation of PbI$_2$ - the following products of perovskite degradation process. Such degradation is accelerated if the sample is stored in ambiant condition as the perovskite layer will be in direct contact with humid environment.  It is noteworthy that PMMA encapsulation not only isolates the perovskite layer from humidity  but also improves the photostability of the perovskite metasurface, thereby reducing the photobleaching effect. Further details of stability studies are shown in Fig.\ref{fig_PMMA} in the Appendix \ref{sec:AppendixPMMA}.

\section{Summary}
To summarize, this work introduced a cost-effective method for fabricating large-scale polariton metasurfaces through direct nano-imprinting of 2D perovskites using a manual hydraulic press. We demonstrated that this straightforward fabrication technique is versatile, applicable to various types of perovskites and different substrates. The strong coupling regime was clearly demonstrated in both reflectivity and photoluminescence experiments, evidenced by the observation of the anti-crossing effect. The measured polaritonic modes align well with theoretical predictions and can exhibit an excitonic fraction of up to 41$\%$ at normal incidence. Impressively, these polaritonic modes are highly homogeneous across centimeter-scale samples, with polariton energy fluctuations measured within a few meV of their linewidth. Additionally, the reproducibility of our method and the stability of the fabricated metasurfaces were thoroughly investigated. Remarkably, it was shown that the same mold could be used more than ten times to create nearly identical metasurfaces. Furthermore, the optical properties of the polariton metasurfaces remained largely unchanged even after 8 months. Such high-quality, cost-effective, large-area metasurfaces hold great promise for future integrated polaritonic devices operating at room temperature.

\textit{Acknowledgement:} The authors would like to thank the staff from the Nanolyon Technical Platform for helping and supporting in all nanofabrication processes. This work is partly supported by the French National Research Agency (ANR) under the project POPEYE (ANR-17-CE24-0020), project EMIPERO (ANR-18-CE24-0016). 
 
\onecolumngrid 
\section{Appendices}
\subsection{Appendix 1: Optical setup}\label{sec:Setup}
The polaritonic dispersions are obtained by angle-resolved micro-reflectivity and micro-photoluminescence using a home-built Fourier spectroscopy setup. Excitation sources include a halogen lamp for reflectivity measurements and a picosecond pulse laser at 405 nm for photoluminescence measurements. The excitation beam is focused onto the sample via a microscope objective ($\times$20, NA=0.42). The reflection/emission signal is collected through the same microscope objective and is then analyzed in the momentum space by imaging the back focal plane of the microscope objective onto the entrance slit of a spectrometer, whose output is coupled to the sensor of a CCD camera.

\subsection{Appendix 2: Supplemental studies on stability}\label{sec:AppendixPMMA}
\begin{figure}[ht!]
	\centering\includegraphics[width=0.5\textwidth]{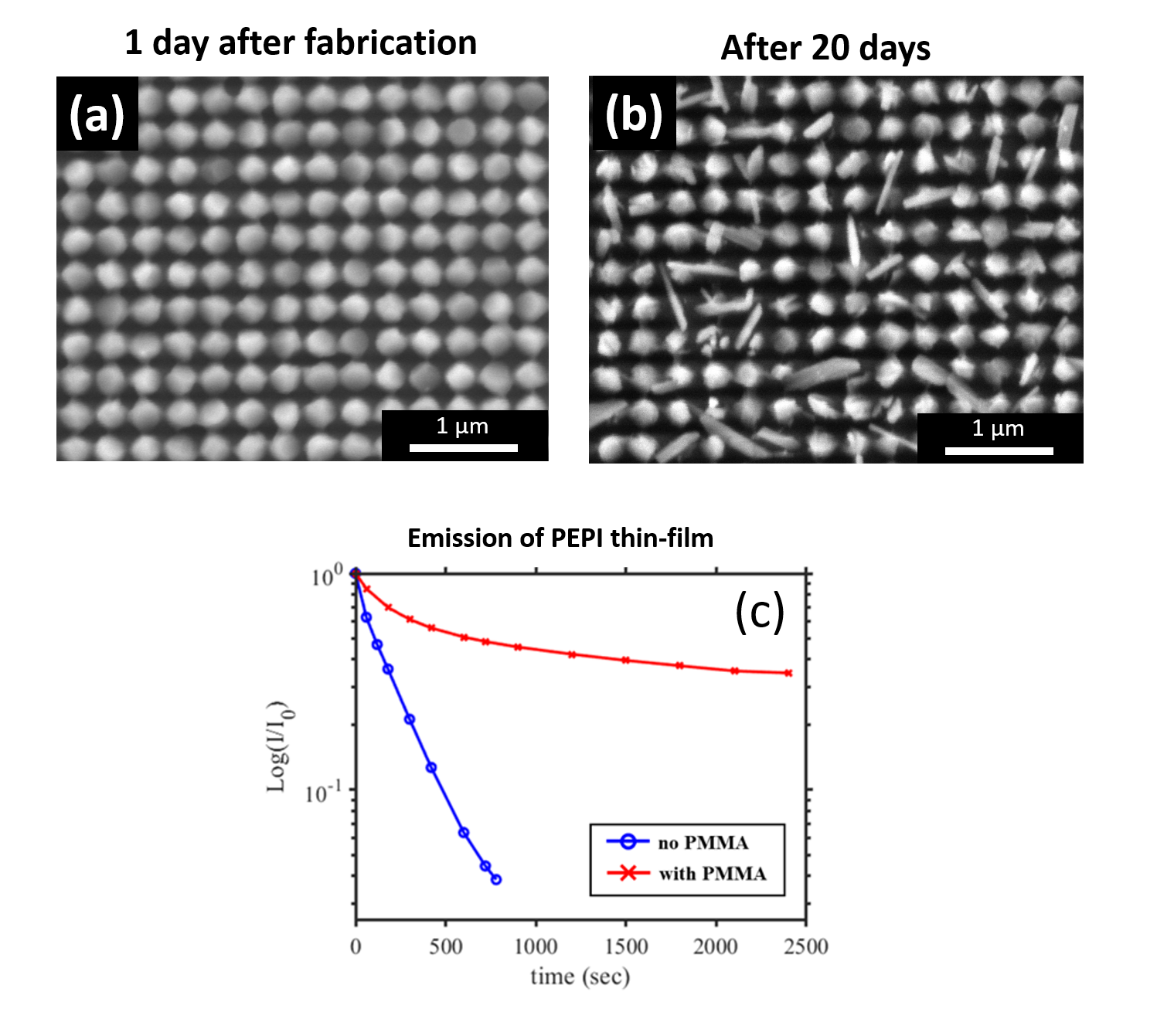}
	\caption{(a,b) SEM images of a BABI metasurface, take one day after fabrication (a) and 20 days after fabrication (b). The sample is not encapsulated with PMMA and is stored under ambient conditions. (c) Degradation of the photoluminescence intensity of a PEPI thin film (with a 10$\%$ weight concentration) deposited on SiO$_2$ substrates is analyzed under two conditions: with PMMA on top (indicated by red crosses) and without PMMA (indicated by blue dots), when continuously exposed to an excitation laser.\label{fig_PMMA}} 
\end{figure}

The degradation of BAPI metasurface is detailed in Figs.\ref{fig_PMMA}a,b, where we observe a significant change in morphology over time with the formation of PbI$_2$, highlighting the material's sensitivity to environmental factors. Additionally, the effectiveness of PMMA encapsulation in enhancing the stability of PEPI thin films is demonstrated in Figs.\ref{fig_PMMA}c. This encapsulation technique not only reduces photobleaching but also contributes to the overall mechanical robustness of the thin films, suggesting its potential for prolonging the operational lifespan of these materials in practical applications. These findings underscore the importance of protective strategies in the development of durable photonic structures.

\subsection{Appendix 3: Theoretical model for strong coupling regime in square lattice metasurface} \label{sec:model}  
\subsubsection{Bloch resonances in square lattice metasurface}
As a first approximation, a periodic metasurface can be regarded as a planar waveguide with the same thickness, and made of a medium with an effective refractive index. In this approach, the effects of periodical corrugation of period $a$ are considered as a perturbation acting on the guided modes of the planar waveguide. By working with a sufficiently thin slab (i.e. subwavelength thickness), and considering only one polarization (for example TE guided modes), the unperturbed planar waveguide can be limited in single mode operation. The folding of this guided mode from the $(n,m)^{th}$ Brillouin zone, given by the Bloch vector $\vec{K}_{n,m}=\frac{2\pi}{a}\left(n \vec{u_x} + m\vec{u_y}\right)$, to the first Brillouin zone corresponds to the folded guided mode $\ket{n,m}$ of wavevector $\vec{k} = k_x \vec{u_x} + k_y \vec{u_y}$. The propagation vector of the folded guided mode $\ket{n,m}$ is given by:
\begin{align}
	\begin{split}
	\vec{\beta}_{n,m}(\vec{k})&=\vec{K}_{n,m} + \vec{k}\\
	& = \left(\frac{2\pi}{a}n+k_x\right) \vec{u_x}  +  \left(\frac{2\pi}{a}m+k_y\right) \vec{u_y}   \label{eq:wavevector_origin}
	\end{split}
\end{align} 
From the effective index $n_e$ and the effective group index $n_g$ of the guided mode in the spectral range of interest, the dispersion characteristic of $\ket{n,m}$ is given by:
\begin{equation}
	\omega_{n,m}(\vec{k})=\frac{c}{n_e}\frac{2\pi}{a} + \frac{c}{n_g}\left(\lvert \vec{\beta}_{n,m}(\vec{k}) \rvert-\frac{2\pi}{a}\right) \label{eq:dispersion_origin}
\end{equation}
where $c$ is the speed of light.
\begin{figure}[htbp]
	\centering\includegraphics[width=0.6\textwidth]{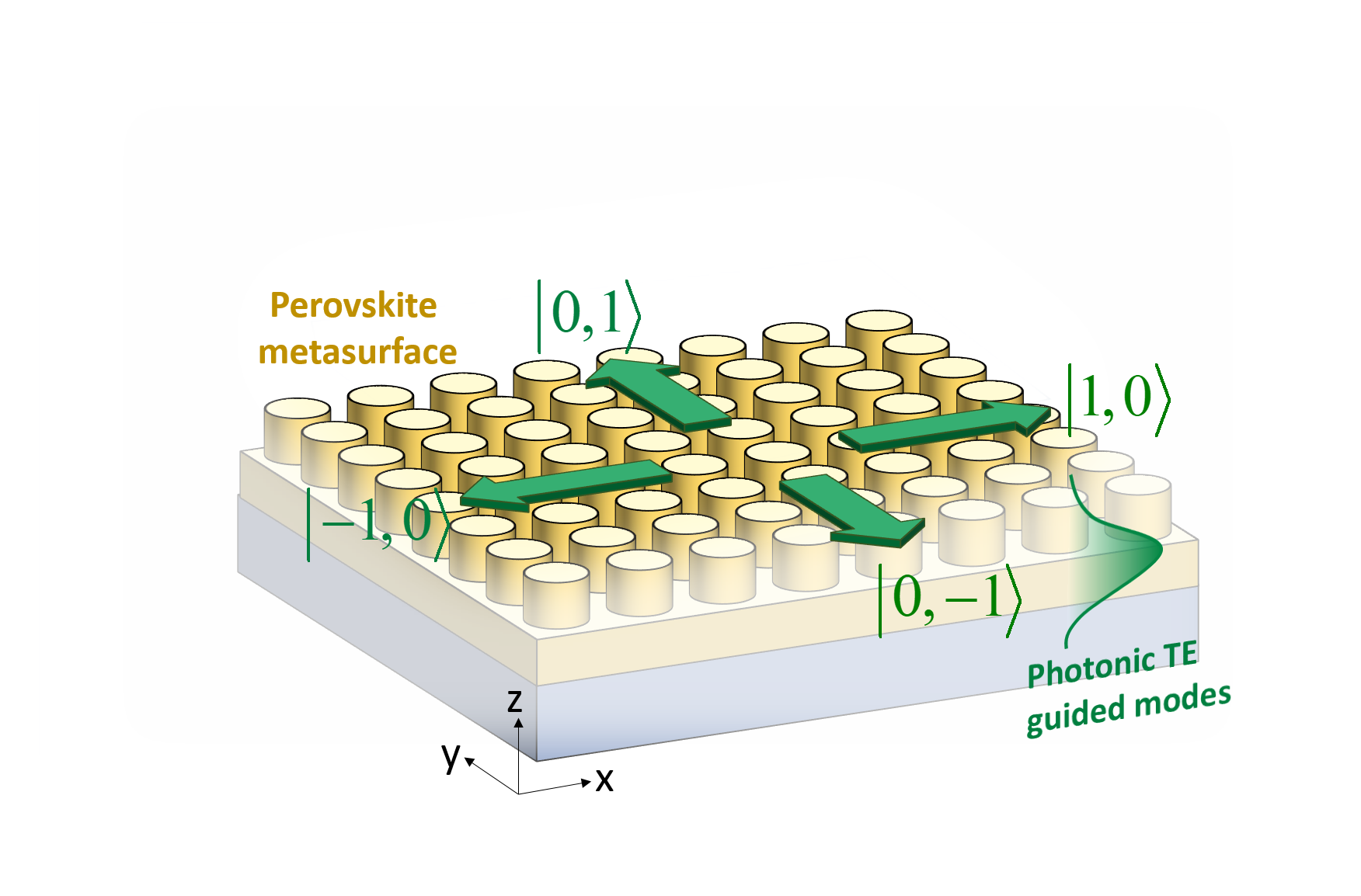}
	\caption{Sketch of the perovsite metasurface. The photonic resonances in this structure are resulted from the coupling between the four folded guided modes $\ket{\pm1,0}$ and $\ket{0,\pm1}$. They are folded from the same TE guided mode of a planar waveguide made of effective index.\label{fig_GMR}} 
\end{figure}

The polarization $\vec{u}_{n,m}$ of the folded guided mode $\ket{n,m}$ can be calculated from its propagation vector $\vec{\beta}_{n,m}$. Indeed, since $\vec{u}_{n,m}.\vec{\beta}_{n,m}=0$ for TE modes, we have:\begin{equation}
	\vec{u}_{n,m}=\frac{- \left(m+ \frac{a k_y}{2\pi}\right)\vec{u_x} + \left(n+ \frac{a k_x}{2\pi}\right)\vec{u_y}}{\sqrt{\left(n+ \frac{a k_x}{2\pi}\right)^2 + \left(m+ \frac{a k_y}{2\pi}\right)^2}}
	\label{eq:u_TE}
\end{equation} 
The period $a$ is chosen so that the spectral range of interest is the one of the first diffraction order, i.e. $|n|+|m|=1$. Therefore we only need to consider four folded guided modes:  $\ket{\pm 1,0}$ and $\ket{0, \pm 1 }$ (see Fig.~\ref{fig_GMR}). Moreover, we only focus at the vicinity of the $\Gamma$ point, thus with $k_{x,y} \ll  2\pi/a$. Thus the dispersion of the folded guided modes are given by:
\begin{align}
	 \begin{split}
	\omega_{\pm 1,0}(\vec{k})&\approx\frac{c}{n_e}\frac{2\pi}{a} \pm \frac{c}{n_g}k_x + \frac{ca}{4\pi n_g}k_y^2,\\ 
	\omega_{0,\pm1}(\vec{k})&\approx\frac{c}{n_e}\frac{2\pi}{a} \pm \frac{c}{n_g}k_y + \frac{ca}{4\pi n_g}k_x^2.
	\label{eq:dispersion_GM}
	 \end{split}
\end{align}
and the corresponding polarizations are:
\begin{align}
	 \begin{split}
	\vec{u}_{\pm 1,0}(\vec{k})&\approx\pm \vec{u_y},\\ 
	\vec{u}_{\pm 0,1}(\vec{k})&\approx\pm \vec{u_x}.
	\label{eq:polarization_GM}
	 \end{split}
\end{align}

The periodic corrugation induces to couplings between folded guided modes via diffractive mechanism, leading to band gap opening at the $\Gamma$ point and hybridization of folded guided modes. In the basis $B= \left\{\ket{1,0},\ket{-1,0},\ket{0,1},\ket{0,-1}\right\}$, the diffractive Hamiltonian is given by $H_d=\begin{psmallmatrix}
		0&U&W&W \\
		U&0&W&W \\
		W&W&0&U \\
		W&W&U&0
	\end{psmallmatrix}$
where $U$ is the coupling strength between counter-propagating modes, and $W$ is the coupling strength between perpendicular-propagating modes. In the following, at first approximation, we assume that $W\ll U$, thus:
\begin{equation}
	H_d\approx\left(\begin{matrix}
		0&U&0&0 \\
		U&0&0&0 \\
		0&0&0&U \\
		0&0&U&0
	\end{matrix}
	\right)
\end{equation}

Moreover, the folded guided modes lie above the light cones, thus radiate to the free space continuum. Such radiative losses is characterized by a coupling strength $\gamma$ corresponding to the radiative linewidth of the guided resonances (here due to the C4 symmetry, the four folded guided modes exhibit the same radiative losses). Ultimately, the folded guided modes can exchange radiative losses via the continuum thanks to Friedrich-Wintgen interference\cite{FW85}. The polarizations given in \eqref{eq:polarization_GM} only allows interference between counter-propagating guided modes. Therefore, in the same basis $B= \left\{\ket{1,0},\ket{-1,0},\ket{0,1},\ket{0,-1}\right\}$,  the radiative Hamiltonian is given by:
\begin{equation}
	H_r=i\gamma\left(\begin{matrix}
		1&-1&0&0 \\
		-1&1&0&0 \\
		0&0&1&-1 \\
		0&0&-1&1
	\end{matrix}
	\right)
\end{equation}
in which the $-1$ coefficients in radiative coupling correspond to the opposite polarizations of counter-propagating waves (see \eqref{eq:polarization_GM}). Therefore, the photoniv Hamiltonian $H_{ph}$ is given by $\begin{psmallmatrix}
		\omega_{1,0}&0&0&0 \\
		0&\omega_{-1,0}&0&0 \\
		0&0&\omega_{0,1}&0 \\
		0&0&0&\omega_{0,-1}
	\end{psmallmatrix} + H_d + H_r$. Finally, considering that we only study the dispersion along $k_x$ (i.e. $k_y=0$), the photonic Hamiltonian is simplified as:
\begin{equation}
	H_{ph}(k_x)=\frac{c}{n_e}\frac{2\pi}{a} + \begin{pmatrix}
		\frac{c}{n_g}k_x + i\gamma&U-i\gamma&0&0 \\
		U-i\gamma&-\frac{c}{n_g}k_x+ i\gamma&0&0 \\
		0&0&\frac{ca}{4n_g\pi}k_x^2+ i\gamma&U-i\gamma\\
		0&0&U-i\gamma&\frac{ca}{4n_g\pi}k_x^2+ i\gamma
	\end{pmatrix}\label{eq:effectiveH}
\end{equation}
The four eigenmodes $\ket{\pm,y}$ and $\ket{\pm,x}$ are of eigenvalues given by:
\begin{align}
	\Omega_{\pm,y}(k_x)&=\frac{c}{n_e}\frac{2\pi}{a} + i\gamma \pm \sqrt{\frac{c^2}{n_g^2}k_x^2 + (U-i\gamma)^2}\\
	\Omega_{\pm,x}(k_x)&=\frac{c}{n_e}\frac{2\pi}{a} + \frac{ca}{4n_g\pi}k_x^2 + i\gamma \pm  (U-i\gamma)
\end{align}
Here, the polarization of $\ket{\pm,y}$ is $\vec{u_y}$ (i.e. $E_y$ polarization), and the one of $\ket{\pm,x}$ is $\vec{u_x}$ (i.e. $E_x$ polarization). At $\Gamma$ point (i.e.$k_x=0$), we have:
\begin{align}
	\Omega_{+,y}(k_x=0)&=\omega_{+,x}(k_x=0)=\frac{c}{n_e}\frac{2\pi}{a} + U\\
	\Omega_{-,y}(k_x=0)&=\omega_{-,x}(k_x=0)=\frac{c}{n_e}\frac{2\pi}{a} - U + 2i\gamma
\end{align}
Therefore $\ket{+,x}$ and $\ket{+,y}$ are lossless at the $\Gamma$ points and become bound states in the continuum (BIC)\cite{Hsu2016}, while  $\ket{-,x}$ and $\ket{-,y}$ take all the losses. We note that $\ket{+,x}$ and $\ket{+,y}$ are non-radiating because their eigenfunction are antisymmetric, thus uncoupled to the radiative continuum due to symmetry mismatch with radiating plane waves. Conversely,  $\ket{-,x}$ and $\ket{-,y}$ exhibit symmetric eigenfunction and can radiate efficiently to the continuum.

In the vicinity of the $\Gamma$ point but at non-zero $k_x$, $\ket{-,x}$ and $\ket{-,y}$ still exhibit much more losses than $\ket{+,x}$ and $\ket{+,y}$. Consequently,  $\ket{+,x}$ and $\ket{+,y}$ are referred to as "dark modes" while $\ket{-,x}$ and $\ket{-,y}$ are referred to as "bright modes".   

\subsubsection{Exciton-photon coupling in square lattice metasurface}
. To study rigorously the formation of exciton-polaritons in subwavelength lattices, a full quantum theory as presented in \cite{Dario2007,Zanotti2022} would be required. Here, we propose a simple model to explain the strong coupling regime involving the four Bloch resonances $\ket{\pm,x}$, $\ket{\pm,y}$, and perovskite excitons. This model is the extension of the one reported in \cite{Lu2020} for 1D polaritonic gratings. As discussed in the previous section, $\ket{\pm,x}$ are of $E_x$ polarization and  $\ket{\pm,y}$ are of $E_y$ polarization. Moreover, the electric fields of $\ket{+,x}$ and $\ket{+,y}$ are antisymmetric, while those of $\ket{-,x}$ and $\ket{-,y}$ are symmetric. Consequently, an exciton with a specific in-plane dipole orientation and location in the perovskite metasurface cannot be efficiently coupled to both photonic modes. As a first approximation, we assume that excitons can be coupled to only one of the four photonic modes. Therefore the polariton Hamiltonian is described by four pairs of
coupled oscillators Hamiltonian, given by:
\begin{equation}
	H_{p}=\begin{pmatrix}
		\omega_X + i\gamma_X & V & 0& 0& 0& 0& 0& 0\\
		V & \Omega_{+,y} & 0& 0& 0& 0& 0& 0 \\
		0& 0&\omega_X + i\gamma_X & V & 0& 0& 0&  0\\
		0& 0& V & \Omega_{-,y} & 0& 0& 0& 0 \\
		0& 0& 0& 0&\omega_X + i\gamma_X & V & 0& 0\\
		0& 0& 0& 0&V & \Omega_{+,x} & 0& 0 \\
		0& 0 & 0& 0& 0&  0 &\omega_X + i\gamma_X& V\\
		0& 0 & 0& 0& 0& 0  &  V & \Omega_{-,x} 
	\end{pmatrix} \label{eq:polaritonH}
\end{equation}
where $\omega_X$, $\gamma_X$ and $V$ correspond to exciton energy, linewidth and the coupling strength with photons respectively. Each pair of coupled oscillators in \eqref{eq:polaritonH} give rises to two polaritonic branches: one upper polariton (UP) and one lower polariton (LP) branch. The nature (i.e. dark/bright and $E_x$/$E_y$ polarization) of the photonic component dictates the nature of the corresponding polaritonic branches.
\subsubsection{Theory vs. Experimental} \label{sec:modelVSexp}
\begin{figure}[htbp]
	\centering\includegraphics[width=0.8\textwidth]{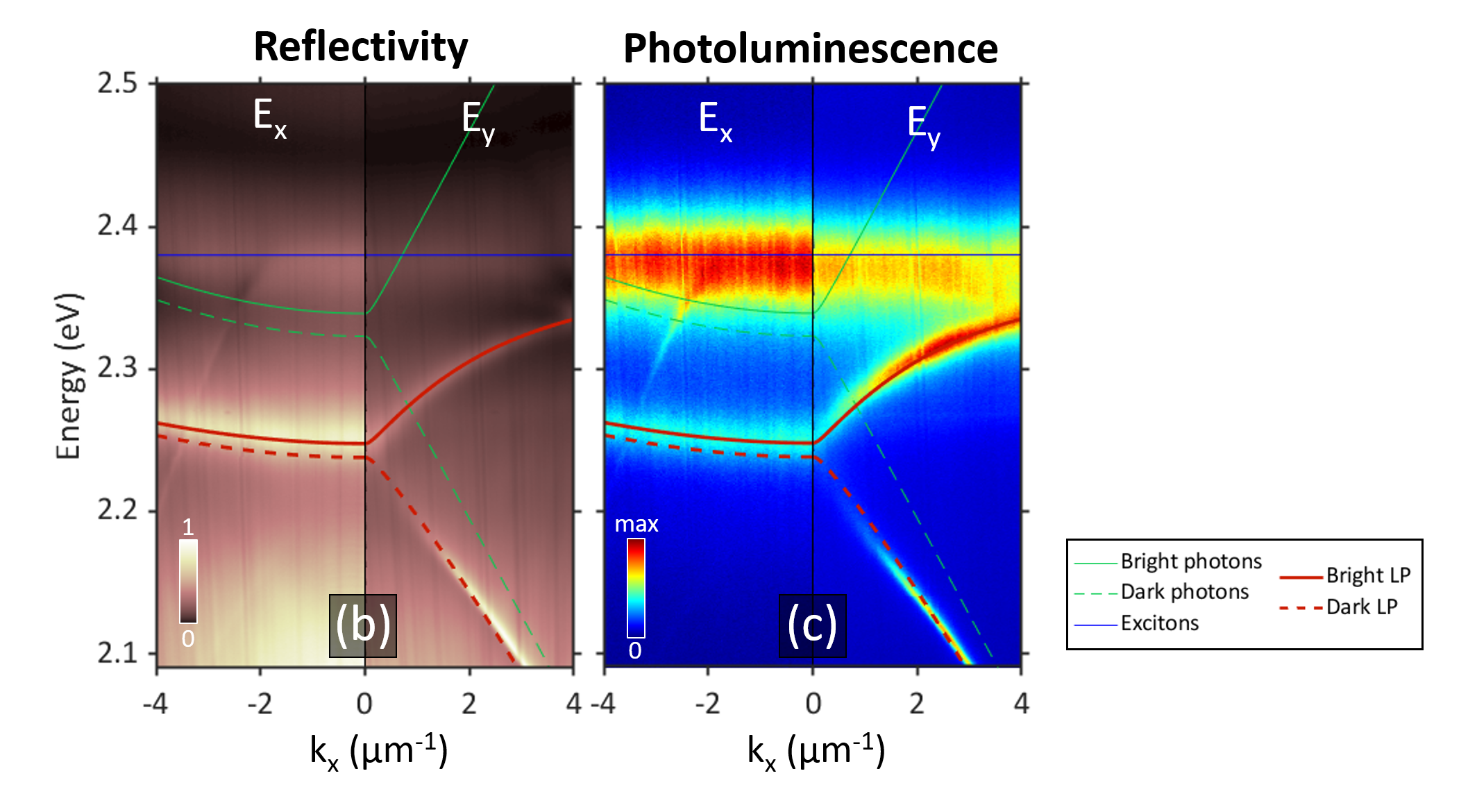}
	\caption{Experimental results of the angle-resolved reflectivity (a) and angle-resolved photoluminescence (b) for the imprinted PEPI layer on quartz substrate. Left panels correspond to the results in $E_x$ polarization, right panels correspond to those in $E_y$ polarization. The fittings obtained from the theoretical model \eqref{eq:polaritonH} are plotted on top of the experimental results. The parameters used in the theoretical model are as follows: $a=300$,nm, $n_e=1.775$, $n_g=2.9$, $\gamma=2$,meV, $U=-8$,meV, $\omega_X=2.38$,eV, $\gamma_X=15$,meV, and $V=110$,meV. .\label{fig_theoryVSexp}} 
\end{figure}

Finally, we apply the theoretical model \eqref{eq:polaritonH} to reproduce the experimental results of polaritonic modes in an imprinted PEPI layer. Measurements in both $E_x$ and $E_y$ polarizations, as reported in Fig.~\ref{fig_theoryVSexp}, show perfect agreement between the experimental results and those obtained from our simple analytical model.. We note that the dispersion characteristic of the polaritonic modes in $E_x$ polarization is almost non-dispersive, thus cannot be used to reveal the anticrossing features of the strong coupling regime..

%%%%%%%%%%%%%%%%%%%%%%% References %%%%%%%%%%%%%%%%%%%%%%%%%

\bibliography{biblio}

\begin{thebibliography}{28}%
\makeatletter
\providecommand \@ifxundefined [1]{%
 \@ifx{#1\undefined}
}%
\providecommand \@ifnum [1]{%
 \ifnum #1\expandafter \@firstoftwo
 \else \expandafter \@secondoftwo
 \fi
}%
\providecommand \@ifx [1]{%
 \ifx #1\expandafter \@firstoftwo
 \else \expandafter \@secondoftwo
 \fi
}%
\providecommand \natexlab [1]{#1}%
\providecommand \enquote  [1]{``#1''}%
\providecommand \bibnamefont  [1]{#1}%
\providecommand \bibfnamefont [1]{#1}%
\providecommand \citenamefont [1]{#1}%
\providecommand \href@noop [0]{\@secondoftwo}%
\providecommand \href [0]{\begingroup \@sanitize@url \@href}%
\providecommand \@href[1]{\@@startlink{#1}\@@href}%
\providecommand \@@href[1]{\endgroup#1\@@endlink}%
\providecommand \@sanitize@url [0]{\catcode `\\12\catcode `\$12\catcode `\&12\catcode `\#12\catcode `\^12\catcode `\_12\catcode `\%12\relax}%
\providecommand \@@startlink[1]{}%
\providecommand \@@endlink[0]{}%
\providecommand \url  [0]{\begingroup\@sanitize@url \@url }%
\providecommand \@url [1]{\endgroup\@href {#1}{\urlprefix }}%
\providecommand \urlprefix  [0]{URL }%
\providecommand \Eprint [0]{\href }%
\providecommand \doibase [0]{https://doi.org/}%
\providecommand \selectlanguage [0]{\@gobble}%
\providecommand \bibinfo  [0]{\@secondoftwo}%
\providecommand \bibfield  [0]{\@secondoftwo}%
\providecommand \translation [1]{[#1]}%
\providecommand \BibitemOpen [0]{}%
\providecommand \bibitemStop [0]{}%
\providecommand \bibitemNoStop [0]{.\EOS\space}%
\providecommand \EOS [0]{\spacefactor3000\relax}%
\providecommand \BibitemShut  [1]{\csname bibitem#1\endcsname}%
\let\auto@bib@innerbib\@empty
%</preamble>
\bibitem [{\citenamefont {Weisbuch}\ \emph {et~al.}(1992)\citenamefont {Weisbuch}, \citenamefont {Nishioka}, \citenamefont {Ishikawa},\ and\ \citenamefont {Arakawa}}]{Claude1994}%
  \BibitemOpen
  \bibfield  {author} {\bibinfo {author} {\bibfnamefont {C.}~\bibnamefont {Weisbuch}}, \bibinfo {author} {\bibfnamefont {M.}~\bibnamefont {Nishioka}}, \bibinfo {author} {\bibfnamefont {A.}~\bibnamefont {Ishikawa}},\ and\ \bibinfo {author} {\bibfnamefont {Y.}~\bibnamefont {Arakawa}},\ }\bibfield  {title} {\bibinfo {title} {Observation of the coupled exciton-photon mode splitting in a semiconductor quantum microcavity},\ }\href {https://doi.org/10.1103/PhysRevLett.69.3314} {\bibfield  {journal} {\bibinfo  {journal} {Phys. Rev. Lett.}\ }\textbf {\bibinfo {volume} {69}},\ \bibinfo {pages} {3314} (\bibinfo {year} {1992})}\BibitemShut {NoStop}%
\bibitem [{\citenamefont {Carusotto}\ and\ \citenamefont {Ciuti}(2013)}]{QuantumFLuidsOfLight}%
  \BibitemOpen
  \bibfield  {author} {\bibinfo {author} {\bibfnamefont {I.}~\bibnamefont {Carusotto}}\ and\ \bibinfo {author} {\bibfnamefont {C.}~\bibnamefont {Ciuti}},\ }\bibfield  {title} {\bibinfo {title} {Quantum fluids of light},\ }\href {https://doi.org/10.1103/RevModPhys.85.299} {\bibfield  {journal} {\bibinfo  {journal} {Rev. Mod. Phys.}\ }\textbf {\bibinfo {volume} {85}},\ \bibinfo {pages} {299} (\bibinfo {year} {2013})}\BibitemShut {NoStop}%
\bibitem [{\citenamefont {Sanvitto}\ and\ \citenamefont {K{\'{e}}na-Cohen}(2016)}]{Sanvitto2016}%
  \BibitemOpen
  \bibfield  {author} {\bibinfo {author} {\bibfnamefont {D.}~\bibnamefont {Sanvitto}}\ and\ \bibinfo {author} {\bibfnamefont {S.}~\bibnamefont {K{\'{e}}na-Cohen}},\ }\bibfield  {title} {\bibinfo {title} {{The road towards polaritonic devices}},\ }\href {https://doi.org/10.1038/nmat4668} {\bibfield  {journal} {\bibinfo  {journal} {Nature Materials}\ }\textbf {\bibinfo {volume} {15}},\ \bibinfo {pages} {doi:10.1038/nmat4668} (\bibinfo {year} {2016})}\BibitemShut {NoStop}%
\bibitem [{\citenamefont {Christopoulos}\ \emph {et~al.}(2007)\citenamefont {Christopoulos}, \citenamefont {{Von H{\"{o}}gersthal}}, \citenamefont {Grundy}, \citenamefont {Lagoudakis}, \citenamefont {Kavokin}, \citenamefont {Baumberg}, \citenamefont {Christmann}, \citenamefont {Butt{\'{e}}}, \citenamefont {Feltin}, \citenamefont {Carlin},\ and\ \citenamefont {Grandjean}}]{Christopoulos2007}%
  \BibitemOpen
  \bibfield  {author} {\bibinfo {author} {\bibfnamefont {S.}~\bibnamefont {Christopoulos}}, \bibinfo {author} {\bibfnamefont {G.~B.~H.}\ \bibnamefont {{Von H{\"{o}}gersthal}}}, \bibinfo {author} {\bibfnamefont {a.~J.~D.}\ \bibnamefont {Grundy}}, \bibinfo {author} {\bibfnamefont {P.~G.}\ \bibnamefont {Lagoudakis}}, \bibinfo {author} {\bibfnamefont {a.~V.}\ \bibnamefont {Kavokin}}, \bibinfo {author} {\bibfnamefont {J.~J.}\ \bibnamefont {Baumberg}}, \bibinfo {author} {\bibfnamefont {G.}~\bibnamefont {Christmann}}, \bibinfo {author} {\bibfnamefont {R.}~\bibnamefont {Butt{\'{e}}}}, \bibinfo {author} {\bibfnamefont {E.}~\bibnamefont {Feltin}}, \bibinfo {author} {\bibfnamefont {J.~F.}\ \bibnamefont {Carlin}},\ and\ \bibinfo {author} {\bibfnamefont {N.}~\bibnamefont {Grandjean}},\ }\bibfield  {title} {\bibinfo {title} {{Room-temperature polariton lasing in semiconductor microcavities}},\ }\href {https://doi.org/10.1103/PhysRevLett.98.126405} {\bibfield  {journal} {\bibinfo  {journal} {Physical Review Letters}\
  }\textbf {\bibinfo {volume} {98}},\ \bibinfo {pages} {1} (\bibinfo {year} {2007})}\BibitemShut {NoStop}%
\bibitem [{\citenamefont {Daskalakis}\ \emph {et~al.}(2013)\citenamefont {Daskalakis}, \citenamefont {Eldridge}, \citenamefont {Christmann}, \citenamefont {Trichas}, \citenamefont {Murray}, \citenamefont {Iliopoulos}, \citenamefont {Monroy}, \citenamefont {Pelekanos}, \citenamefont {Baumberg},\ and\ \citenamefont {Savvidis}}]{Daskalakis2013}%
  \BibitemOpen
  \bibfield  {author} {\bibinfo {author} {\bibfnamefont {K.~S.}\ \bibnamefont {Daskalakis}}, \bibinfo {author} {\bibfnamefont {P.~S.}\ \bibnamefont {Eldridge}}, \bibinfo {author} {\bibfnamefont {G.}~\bibnamefont {Christmann}}, \bibinfo {author} {\bibfnamefont {E.}~\bibnamefont {Trichas}}, \bibinfo {author} {\bibfnamefont {R.}~\bibnamefont {Murray}}, \bibinfo {author} {\bibfnamefont {E.}~\bibnamefont {Iliopoulos}}, \bibinfo {author} {\bibfnamefont {E.}~\bibnamefont {Monroy}}, \bibinfo {author} {\bibfnamefont {N.~T.}\ \bibnamefont {Pelekanos}}, \bibinfo {author} {\bibfnamefont {J.~J.}\ \bibnamefont {Baumberg}},\ and\ \bibinfo {author} {\bibfnamefont {P.~G.}\ \bibnamefont {Savvidis}},\ }\bibfield  {title} {\bibinfo {title} {{All-dielectric GaN microcavity: Strong coupling and lasing at room temperature}},\ }\bibfield  {journal} {\bibinfo  {journal} {Applied Physics Letters}\ }\textbf {\bibinfo {volume} {102}},\ \href {https://doi.org/10.1063/1.4795019} {10.1063/1.4795019} (\bibinfo {year} {2013})\BibitemShut
  {NoStop}%
\bibitem [{\citenamefont {Franke}\ \emph {et~al.}(2012)\citenamefont {Franke}, \citenamefont {Sturm}, \citenamefont {Schmidt-Grund}, \citenamefont {Wagner},\ and\ \citenamefont {Grundmann}}]{Franke_2012}%
  \BibitemOpen
  \bibfield  {author} {\bibinfo {author} {\bibfnamefont {H.}~\bibnamefont {Franke}}, \bibinfo {author} {\bibfnamefont {C.}~\bibnamefont {Sturm}}, \bibinfo {author} {\bibfnamefont {R.}~\bibnamefont {Schmidt-Grund}}, \bibinfo {author} {\bibfnamefont {G.}~\bibnamefont {Wagner}},\ and\ \bibinfo {author} {\bibfnamefont {M.}~\bibnamefont {Grundmann}},\ }\bibfield  {title} {\bibinfo {title} {{Ballistic propagation of exciton polariton condensates in a ZnO-based microcavity}},\ }\href {https://doi.org/10.1088/1367-2630/14/1/013037} {\bibfield  {journal} {\bibinfo  {journal} {New Journal of Physics}\ }\textbf {\bibinfo {volume} {14}},\ \bibinfo {pages} {13037} (\bibinfo {year} {2012})}\BibitemShut {NoStop}%
\bibitem [{\citenamefont {Lidzey}\ \emph {et~al.}(1998)\citenamefont {Lidzey}, \citenamefont {Bradley}, \citenamefont {Skolnick}, \citenamefont {Virgili}, \citenamefont {Walker},\ and\ \citenamefont {Whittaker}}]{Lidzey1998}%
  \BibitemOpen
  \bibfield  {author} {\bibinfo {author} {\bibfnamefont {D.~G.}\ \bibnamefont {Lidzey}}, \bibinfo {author} {\bibfnamefont {D.~D.~C.}\ \bibnamefont {Bradley}}, \bibinfo {author} {\bibfnamefont {M.~S.}\ \bibnamefont {Skolnick}}, \bibinfo {author} {\bibfnamefont {T.}~\bibnamefont {Virgili}}, \bibinfo {author} {\bibfnamefont {S.}~\bibnamefont {Walker}},\ and\ \bibinfo {author} {\bibfnamefont {D.~M.}\ \bibnamefont {Whittaker}},\ }\bibfield  {title} {\bibinfo {title} {{Strong exciton–photon coupling in an organic semiconductor microcavity}},\ }\href {https://doi.org/10.1038/25692} {\bibfield  {journal} {\bibinfo  {journal} {Nature}\ }\textbf {\bibinfo {volume} {395}},\ \bibinfo {pages} {53} (\bibinfo {year} {1998})}\BibitemShut {NoStop}%
\bibitem [{\citenamefont {Plumhof}\ \emph {et~al.}(2014)\citenamefont {Plumhof}, \citenamefont {St{\"{o}}ferle}, \citenamefont {Mai}, \citenamefont {Scherf},\ and\ \citenamefont {Mahrt}}]{Plumhof2014}%
  \BibitemOpen
  \bibfield  {author} {\bibinfo {author} {\bibfnamefont {J.~D.}\ \bibnamefont {Plumhof}}, \bibinfo {author} {\bibfnamefont {T.}~\bibnamefont {St{\"{o}}ferle}}, \bibinfo {author} {\bibfnamefont {L.}~\bibnamefont {Mai}}, \bibinfo {author} {\bibfnamefont {U.}~\bibnamefont {Scherf}},\ and\ \bibinfo {author} {\bibfnamefont {R.~F.}\ \bibnamefont {Mahrt}},\ }\bibfield  {title} {\bibinfo {title} {{Room-temperature Bose–Einstein condensation of cavity exciton–polaritons in a polymer}},\ }\href {https://doi.org/10.1038/nmat3825} {\bibfield  {journal} {\bibinfo  {journal} {Nature Materials}\ }\textbf {\bibinfo {volume} {13}},\ \bibinfo {pages} {247} (\bibinfo {year} {2014})}\BibitemShut {NoStop}%
\bibitem [{\citenamefont {Liu}\ \emph {et~al.}(2014)\citenamefont {Liu}, \citenamefont {Galfsky}, \citenamefont {Sun}, \citenamefont {Xia}, \citenamefont {Lin}, \citenamefont {Lee}, \citenamefont {K{\'{e}}na-Cohen},\ and\ \citenamefont {Menon}}]{Liu2014}%
  \BibitemOpen
  \bibfield  {author} {\bibinfo {author} {\bibfnamefont {X.}~\bibnamefont {Liu}}, \bibinfo {author} {\bibfnamefont {T.}~\bibnamefont {Galfsky}}, \bibinfo {author} {\bibfnamefont {Z.}~\bibnamefont {Sun}}, \bibinfo {author} {\bibfnamefont {F.}~\bibnamefont {Xia}}, \bibinfo {author} {\bibfnamefont {E.-c.}\ \bibnamefont {Lin}}, \bibinfo {author} {\bibfnamefont {Y.-H.}\ \bibnamefont {Lee}}, \bibinfo {author} {\bibfnamefont {S.}~\bibnamefont {K{\'{e}}na-Cohen}},\ and\ \bibinfo {author} {\bibfnamefont {V.~M.}\ \bibnamefont {Menon}},\ }\bibfield  {title} {\bibinfo {title} {{Strong light–matter coupling in two-dimensional atomic crystals}},\ }\href {https://doi.org/10.1038/nphoton.2014.304 http://10.0.4.14/nphoton.2014.304 https://www.nature.com/articles/nphoton.2014.304{\#}supplementary-information} {\bibfield  {journal} {\bibinfo  {journal} {Nature Photonics}\ }\textbf {\bibinfo {volume} {9}},\ \bibinfo {pages} {30} (\bibinfo {year} {2014})}\BibitemShut {NoStop}%
\bibitem [{\citenamefont {Grosso}(2017)}]{Grosso2017}%
  \BibitemOpen
  \bibfield  {author} {\bibinfo {author} {\bibfnamefont {G.}~\bibnamefont {Grosso}},\ }\bibfield  {title} {\bibinfo {title} {{2D materials: Valley polaritons}},\ }\href {https://doi.org/10.1038/nphoton.2017.135} {\bibfield  {journal} {\bibinfo  {journal} {Nature Photonics}\ }\textbf {\bibinfo {volume} {11}},\ \bibinfo {pages} {455} (\bibinfo {year} {2017})}\BibitemShut {NoStop}%
\bibitem [{\citenamefont {Su}\ \emph {et~al.}(2021)\citenamefont {Su}, \citenamefont {Fieramosca}, \citenamefont {Zhang}, \citenamefont {Nguyen}, \citenamefont {Deleporte}, \citenamefont {Chen}, \citenamefont {Sanvitto}, \citenamefont {Liew},\ and\ \citenamefont {Xiong}}]{Su2021}%
  \BibitemOpen
  \bibfield  {author} {\bibinfo {author} {\bibfnamefont {R.}~\bibnamefont {Su}}, \bibinfo {author} {\bibfnamefont {A.}~\bibnamefont {Fieramosca}}, \bibinfo {author} {\bibfnamefont {Q.}~\bibnamefont {Zhang}}, \bibinfo {author} {\bibfnamefont {H.~S.}\ \bibnamefont {Nguyen}}, \bibinfo {author} {\bibfnamefont {E.}~\bibnamefont {Deleporte}}, \bibinfo {author} {\bibfnamefont {Z.}~\bibnamefont {Chen}}, \bibinfo {author} {\bibfnamefont {D.}~\bibnamefont {Sanvitto}}, \bibinfo {author} {\bibfnamefont {T.~C.~H.}\ \bibnamefont {Liew}},\ and\ \bibinfo {author} {\bibfnamefont {Q.}~\bibnamefont {Xiong}},\ }\bibfield  {title} {\bibinfo {title} {{Perovskite semiconductors for room-temperature exciton-polaritonics}},\ }\bibfield  {journal} {\bibinfo  {journal} {Nature Materials}\ }\href {https://doi.org/10.1038/s41563-021-01035-x} {10.1038/s41563-021-01035-x} (\bibinfo {year} {2021})\BibitemShut {NoStop}%
\bibitem [{\citenamefont {Dang}\ \emph {et~al.}(2020)\citenamefont {Dang}, \citenamefont {Gerace}, \citenamefont {Drouard}, \citenamefont {Trippé-Allard}, \citenamefont {Lédée}, \citenamefont {Mazurczyk}, \citenamefont {Deleporte}, \citenamefont {Seassal},\ and\ \citenamefont {Nguyen}}]{Dang2020}%
  \BibitemOpen
  \bibfield  {author} {\bibinfo {author} {\bibfnamefont {N.~H.~M.}\ \bibnamefont {Dang}}, \bibinfo {author} {\bibfnamefont {D.}~\bibnamefont {Gerace}}, \bibinfo {author} {\bibfnamefont {E.}~\bibnamefont {Drouard}}, \bibinfo {author} {\bibfnamefont {G.}~\bibnamefont {Trippé-Allard}}, \bibinfo {author} {\bibfnamefont {F.}~\bibnamefont {Lédée}}, \bibinfo {author} {\bibfnamefont {R.}~\bibnamefont {Mazurczyk}}, \bibinfo {author} {\bibfnamefont {E.}~\bibnamefont {Deleporte}}, \bibinfo {author} {\bibfnamefont {C.}~\bibnamefont {Seassal}},\ and\ \bibinfo {author} {\bibfnamefont {H.~S.}\ \bibnamefont {Nguyen}},\ }\bibfield  {title} {\bibinfo {title} {Tailoring dispersion of room-temperature exciton-polaritons with perovskite-based subwavelength metasurfaces},\ }\href {https://doi.org/10.1021/acs.nanolett.0c00125} {\bibfield  {journal} {\bibinfo  {journal} {Nano Lett.}\ }\textbf {\bibinfo {volume} {20}},\ \bibinfo {pages} {2113} (\bibinfo {year} {2020})}\BibitemShut {NoStop}%
\bibitem [{\citenamefont {Lu}\ \emph {et~al.}(2020)\citenamefont {Lu}, \citenamefont {Le-Van}, \citenamefont {Ferrier}, \citenamefont {Drouard}, \citenamefont {Seassal},\ and\ \citenamefont {Nguyen}}]{Lu2020}%
  \BibitemOpen
  \bibfield  {author} {\bibinfo {author} {\bibfnamefont {L.}~\bibnamefont {Lu}}, \bibinfo {author} {\bibfnamefont {Q.}~\bibnamefont {Le-Van}}, \bibinfo {author} {\bibfnamefont {L.}~\bibnamefont {Ferrier}}, \bibinfo {author} {\bibfnamefont {E.}~\bibnamefont {Drouard}}, \bibinfo {author} {\bibfnamefont {C.}~\bibnamefont {Seassal}},\ and\ \bibinfo {author} {\bibfnamefont {H.~S.}\ \bibnamefont {Nguyen}},\ }\bibfield  {title} {\bibinfo {title} {Engineering a light-matter strong coupling regime in perovskite-based plasmonic metasurface: quasi-bound state in the continuum and exceptional points},\ }\href {https://doi.org/10.1364/PRJ.404743} {\bibfield  {journal} {\bibinfo  {journal} {Photon. Res.}\ }\textbf {\bibinfo {volume} {8}},\ \bibinfo {pages} {A91} (\bibinfo {year} {2020})}\BibitemShut {NoStop}%
\bibitem [{\citenamefont {Dang}\ \emph {et~al.}(2022)\citenamefont {Dang}, \citenamefont {Zanotti}, \citenamefont {Drouard}, \citenamefont {Chevalier}, \citenamefont {Trippé-Allard}, \citenamefont {Amara}, \citenamefont {Deleporte}, \citenamefont {Ardizzone}, \citenamefont {Sanvitto}, \citenamefont {Andreani}, \citenamefont {Seassal}, \citenamefont {Gerace},\ and\ \citenamefont {Nguyen}}]{Dang2022}%
  \BibitemOpen
  \bibfield  {author} {\bibinfo {author} {\bibfnamefont {N.~H.~M.}\ \bibnamefont {Dang}}, \bibinfo {author} {\bibfnamefont {S.}~\bibnamefont {Zanotti}}, \bibinfo {author} {\bibfnamefont {E.}~\bibnamefont {Drouard}}, \bibinfo {author} {\bibfnamefont {C.}~\bibnamefont {Chevalier}}, \bibinfo {author} {\bibfnamefont {G.}~\bibnamefont {Trippé-Allard}}, \bibinfo {author} {\bibfnamefont {M.}~\bibnamefont {Amara}}, \bibinfo {author} {\bibfnamefont {E.}~\bibnamefont {Deleporte}}, \bibinfo {author} {\bibfnamefont {V.}~\bibnamefont {Ardizzone}}, \bibinfo {author} {\bibfnamefont {D.}~\bibnamefont {Sanvitto}}, \bibinfo {author} {\bibfnamefont {L.~C.}\ \bibnamefont {Andreani}}, \bibinfo {author} {\bibfnamefont {C.}~\bibnamefont {Seassal}}, \bibinfo {author} {\bibfnamefont {D.}~\bibnamefont {Gerace}},\ and\ \bibinfo {author} {\bibfnamefont {H.~S.}\ \bibnamefont {Nguyen}},\ }\bibfield  {title} {\bibinfo {title} {Realization of polaritonic topological charge at room temperature using polariton bound states in the continuum
  from perovskite metasurface},\ }\href {https://doi.org/https://doi.org/10.1002/adom.202102386} {\bibfield  {journal} {\bibinfo  {journal} {Advanced Optical Materials}\ }\textbf {\bibinfo {volume} {10}},\ \bibinfo {pages} {2102386} (\bibinfo {year} {2022})}\BibitemShut {NoStop}%
\bibitem [{\citenamefont {Kim}\ \emph {et~al.}(2021)\citenamefont {Kim}, \citenamefont {Woo}, \citenamefont {An}, \citenamefont {Lim}, \citenamefont {Seo}, \citenamefont {Kim}, \citenamefont {Yoo}, \citenamefont {Park},\ and\ \citenamefont {Jun}}]{Kim2021}%
  \BibitemOpen
  \bibfield  {author} {\bibinfo {author} {\bibfnamefont {S.}~\bibnamefont {Kim}}, \bibinfo {author} {\bibfnamefont {B.~H.}\ \bibnamefont {Woo}}, \bibinfo {author} {\bibfnamefont {S.-C.}\ \bibnamefont {An}}, \bibinfo {author} {\bibfnamefont {Y.}~\bibnamefont {Lim}}, \bibinfo {author} {\bibfnamefont {I.~C.}\ \bibnamefont {Seo}}, \bibinfo {author} {\bibfnamefont {D.-S.}\ \bibnamefont {Kim}}, \bibinfo {author} {\bibfnamefont {S.}~\bibnamefont {Yoo}}, \bibinfo {author} {\bibfnamefont {Q.-H.}\ \bibnamefont {Park}},\ and\ \bibinfo {author} {\bibfnamefont {Y.~C.}\ \bibnamefont {Jun}},\ }\bibfield  {title} {\bibinfo {title} {{Topological Control of 2D Perovskite Emission in the Strong Coupling Regime}},\ }\href {https://doi.org/10.1021/acs.nanolett.1c03853} {\bibfield  {journal} {\bibinfo  {journal} {Nano Letters}\ }\textbf {\bibinfo {volume} {21}},\ \bibinfo {pages} {10076} (\bibinfo {year} {2021})}\BibitemShut {NoStop}%
\bibitem [{\citenamefont {Ardizzone}\ \emph {et~al.}(2022)\citenamefont {Ardizzone}, \citenamefont {Riminucci}, \citenamefont {Zanotti}, \citenamefont {Gianfrate}, \citenamefont {Efthymiou-Tsironi}, \citenamefont {Su{\`a}rez-Forero}, \citenamefont {Todisco}, \citenamefont {De~Giorgi}, \citenamefont {Trypogeorgos}, \citenamefont {Gigli}, \citenamefont {Baldwin}, \citenamefont {Pfeiffer}, \citenamefont {Ballarini}, \citenamefont {Nguyen}, \citenamefont {Gerace},\ and\ \citenamefont {Sanvitto}}]{Ardizzone_Nature2022}%
  \BibitemOpen
  \bibfield  {author} {\bibinfo {author} {\bibfnamefont {V.}~\bibnamefont {Ardizzone}}, \bibinfo {author} {\bibfnamefont {F.}~\bibnamefont {Riminucci}}, \bibinfo {author} {\bibfnamefont {S.}~\bibnamefont {Zanotti}}, \bibinfo {author} {\bibfnamefont {A.}~\bibnamefont {Gianfrate}}, \bibinfo {author} {\bibfnamefont {M.}~\bibnamefont {Efthymiou-Tsironi}}, \bibinfo {author} {\bibfnamefont {D.~G.}\ \bibnamefont {Su{\`a}rez-Forero}}, \bibinfo {author} {\bibfnamefont {F.}~\bibnamefont {Todisco}}, \bibinfo {author} {\bibfnamefont {M.}~\bibnamefont {De~Giorgi}}, \bibinfo {author} {\bibfnamefont {D.}~\bibnamefont {Trypogeorgos}}, \bibinfo {author} {\bibfnamefont {G.}~\bibnamefont {Gigli}}, \bibinfo {author} {\bibfnamefont {K.}~\bibnamefont {Baldwin}}, \bibinfo {author} {\bibfnamefont {L.}~\bibnamefont {Pfeiffer}}, \bibinfo {author} {\bibfnamefont {D.}~\bibnamefont {Ballarini}}, \bibinfo {author} {\bibfnamefont {H.~S.}\ \bibnamefont {Nguyen}}, \bibinfo {author} {\bibfnamefont {D.}~\bibnamefont {Gerace}},\ and\ \bibinfo
  {author} {\bibfnamefont {D.}~\bibnamefont {Sanvitto}},\ }\bibfield  {title} {\bibinfo {title} {Polariton {B}ose--{E}instein condensate from a bound state in the continuum},\ }\href {https://doi.org/10.1038/s41586-022-04583-7} {\bibfield  {journal} {\bibinfo  {journal} {Nature}\ }\textbf {\bibinfo {volume} {605}},\ \bibinfo {pages} {447} (\bibinfo {year} {2022})}\BibitemShut {NoStop}%
\bibitem [{\citenamefont {Gianfrate}\ \emph {et~al.}(2023)\citenamefont {Gianfrate}, \citenamefont {Sigurdsson}, \citenamefont {Ardizzone}, \citenamefont {Nguyen}, \citenamefont {Riminucci}, \citenamefont {Efthymiou-Tsironi}, \citenamefont {Baldwin}, \citenamefont {Pfeiffer}, \citenamefont {Trypogeorgos}, \citenamefont {Giorgi}, \citenamefont {Ballarini}, \citenamefont {Nguyen},\ and\ \citenamefont {Sanvitto}}]{gianfrate2023optically}%
  \BibitemOpen
  \bibfield  {author} {\bibinfo {author} {\bibfnamefont {A.}~\bibnamefont {Gianfrate}}, \bibinfo {author} {\bibfnamefont {H.}~\bibnamefont {Sigurdsson}}, \bibinfo {author} {\bibfnamefont {V.}~\bibnamefont {Ardizzone}}, \bibinfo {author} {\bibfnamefont {H.~C.}\ \bibnamefont {Nguyen}}, \bibinfo {author} {\bibfnamefont {F.}~\bibnamefont {Riminucci}}, \bibinfo {author} {\bibfnamefont {M.}~\bibnamefont {Efthymiou-Tsironi}}, \bibinfo {author} {\bibfnamefont {K.~W.}\ \bibnamefont {Baldwin}}, \bibinfo {author} {\bibfnamefont {L.~N.}\ \bibnamefont {Pfeiffer}}, \bibinfo {author} {\bibfnamefont {D.}~\bibnamefont {Trypogeorgos}}, \bibinfo {author} {\bibfnamefont {M.~D.}\ \bibnamefont {Giorgi}}, \bibinfo {author} {\bibfnamefont {D.}~\bibnamefont {Ballarini}}, \bibinfo {author} {\bibfnamefont {H.~S.}\ \bibnamefont {Nguyen}},\ and\ \bibinfo {author} {\bibfnamefont {D.}~\bibnamefont {Sanvitto}},\ }\href@noop {} {\bibinfo {title} {Optically reconfigurable molecules of topological bound states in the continuum}} (\bibinfo
  {year} {2023}),\ \Eprint {https://arxiv.org/abs/2301.08477} {arXiv:2301.08477 [physics.optics]} \BibitemShut {NoStop}%
\bibitem [{\citenamefont {Riminucci}\ \emph {et~al.}(2023)\citenamefont {Riminucci}, \citenamefont {Gianfrate}, \citenamefont {Nigro}, \citenamefont {Ardizzone}, \citenamefont {Dhuey}, \citenamefont {Francaviglia}, \citenamefont {Baldwin}, \citenamefont {Pfeiffer}, \citenamefont {Trypogeorgos}, \citenamefont {Schwartzberg}, \citenamefont {Gerace},\ and\ \citenamefont {Sanvitto}}]{riminucci2023boseeinstein}%
  \BibitemOpen
  \bibfield  {author} {\bibinfo {author} {\bibfnamefont {F.}~\bibnamefont {Riminucci}}, \bibinfo {author} {\bibfnamefont {A.}~\bibnamefont {Gianfrate}}, \bibinfo {author} {\bibfnamefont {D.}~\bibnamefont {Nigro}}, \bibinfo {author} {\bibfnamefont {V.}~\bibnamefont {Ardizzone}}, \bibinfo {author} {\bibfnamefont {S.}~\bibnamefont {Dhuey}}, \bibinfo {author} {\bibfnamefont {L.}~\bibnamefont {Francaviglia}}, \bibinfo {author} {\bibfnamefont {K.}~\bibnamefont {Baldwin}}, \bibinfo {author} {\bibfnamefont {L.~N.}\ \bibnamefont {Pfeiffer}}, \bibinfo {author} {\bibfnamefont {D.}~\bibnamefont {Trypogeorgos}}, \bibinfo {author} {\bibfnamefont {A.}~\bibnamefont {Schwartzberg}}, \bibinfo {author} {\bibfnamefont {D.}~\bibnamefont {Gerace}},\ and\ \bibinfo {author} {\bibfnamefont {D.}~\bibnamefont {Sanvitto}},\ }\href@noop {} {\bibinfo {title} {Bose-einstein condensation in gap-confined exciton-polariton states}} (\bibinfo {year} {2023}),\ \Eprint {https://arxiv.org/abs/2306.03309} {arXiv:2306.03309 [cond-mat.mes-hall]}
  \BibitemShut {NoStop}%
\bibitem [{\citenamefont {Berghuis}\ \emph {et~al.}(2023)\citenamefont {Berghuis}, \citenamefont {Castellanos}, \citenamefont {Murai}, \citenamefont {Pura}, \citenamefont {Abujetas}, \citenamefont {van Heijst}, \citenamefont {Ramezani}, \citenamefont {S{\'{a}}nchez-Gil},\ and\ \citenamefont {Rivas}}]{Berghuis2023}%
  \BibitemOpen
  \bibfield  {author} {\bibinfo {author} {\bibfnamefont {A.~M.}\ \bibnamefont {Berghuis}}, \bibinfo {author} {\bibfnamefont {G.~W.}\ \bibnamefont {Castellanos}}, \bibinfo {author} {\bibfnamefont {S.}~\bibnamefont {Murai}}, \bibinfo {author} {\bibfnamefont {J.~L.}\ \bibnamefont {Pura}}, \bibinfo {author} {\bibfnamefont {D.~R.}\ \bibnamefont {Abujetas}}, \bibinfo {author} {\bibfnamefont {E.}~\bibnamefont {van Heijst}}, \bibinfo {author} {\bibfnamefont {M.}~\bibnamefont {Ramezani}}, \bibinfo {author} {\bibfnamefont {J.~A.}\ \bibnamefont {S{\'{a}}nchez-Gil}},\ and\ \bibinfo {author} {\bibfnamefont {J.~G.}\ \bibnamefont {Rivas}},\ }\bibfield  {title} {\bibinfo {title} {{Room Temperature Exciton–Polariton Condensation in Silicon Metasurfaces Emerging from Bound States in the Continuum}},\ }\href {https://doi.org/10.1021/acs.nanolett.3c01102} {\bibfield  {journal} {\bibinfo  {journal} {Nano Letters}\ }\textbf {\bibinfo {volume} {23}},\ \bibinfo {pages} {5603} (\bibinfo {year} {2023})}\BibitemShut {NoStop}%
\bibitem [{\citenamefont {Kravtsov}\ \emph {et~al.}(2020)\citenamefont {Kravtsov}, \citenamefont {Khestanova}, \citenamefont {Benimetskiy}, \citenamefont {Ivanova}, \citenamefont {Samusev}, \citenamefont {Sinev}, \citenamefont {Pidgayko}, \citenamefont {Mozharov}, \citenamefont {Mukhin}, \citenamefont {Lozhkin}, \citenamefont {Kapitonov}, \citenamefont {Brichkin}, \citenamefont {Kulakovskii}, \citenamefont {Shelykh}, \citenamefont {Tartakovskii}, \citenamefont {Walker}, \citenamefont {Skolnick}, \citenamefont {Krizhanovskii},\ and\ \citenamefont {Iorsh}}]{Kravtsov2020}%
  \BibitemOpen
  \bibfield  {author} {\bibinfo {author} {\bibfnamefont {V.}~\bibnamefont {Kravtsov}}, \bibinfo {author} {\bibfnamefont {E.}~\bibnamefont {Khestanova}}, \bibinfo {author} {\bibfnamefont {F.~A.}\ \bibnamefont {Benimetskiy}}, \bibinfo {author} {\bibfnamefont {T.}~\bibnamefont {Ivanova}}, \bibinfo {author} {\bibfnamefont {A.~K.}\ \bibnamefont {Samusev}}, \bibinfo {author} {\bibfnamefont {I.~S.}\ \bibnamefont {Sinev}}, \bibinfo {author} {\bibfnamefont {D.}~\bibnamefont {Pidgayko}}, \bibinfo {author} {\bibfnamefont {A.~M.}\ \bibnamefont {Mozharov}}, \bibinfo {author} {\bibfnamefont {I.~S.}\ \bibnamefont {Mukhin}}, \bibinfo {author} {\bibfnamefont {M.~S.}\ \bibnamefont {Lozhkin}}, \bibinfo {author} {\bibfnamefont {Y.~V.}\ \bibnamefont {Kapitonov}}, \bibinfo {author} {\bibfnamefont {A.~S.}\ \bibnamefont {Brichkin}}, \bibinfo {author} {\bibfnamefont {V.~D.}\ \bibnamefont {Kulakovskii}}, \bibinfo {author} {\bibfnamefont {I.~A.}\ \bibnamefont {Shelykh}}, \bibinfo {author} {\bibfnamefont {A.~I.}\ \bibnamefont
  {Tartakovskii}}, \bibinfo {author} {\bibfnamefont {P.~M.}\ \bibnamefont {Walker}}, \bibinfo {author} {\bibfnamefont {M.~S.}\ \bibnamefont {Skolnick}}, \bibinfo {author} {\bibfnamefont {D.~N.}\ \bibnamefont {Krizhanovskii}},\ and\ \bibinfo {author} {\bibfnamefont {I.~V.}\ \bibnamefont {Iorsh}},\ }\bibfield  {title} {\bibinfo {title} {{Nonlinear polaritons in a monolayer semiconductor coupled to optical bound states in the continuum}},\ }\href {https://doi.org/10.1038/s41377-020-0286-z} {\bibfield  {journal} {\bibinfo  {journal} {Light: Science \& Applications}\ }\textbf {\bibinfo {volume} {9}},\ \bibinfo {pages} {56} (\bibinfo {year} {2020})}\BibitemShut {NoStop}%
\bibitem [{\citenamefont {Maggiolini}\ \emph {et~al.}(2023)\citenamefont {Maggiolini}, \citenamefont {Polimeno}, \citenamefont {Todisco}, \citenamefont {{Di Renzo}}, \citenamefont {Han}, \citenamefont {{De Giorgi}}, \citenamefont {Ardizzone}, \citenamefont {Schneider}, \citenamefont {Mastria}, \citenamefont {Cannavale}, \citenamefont {Pugliese}, \citenamefont {{De Marco}}, \citenamefont {Rizzo}, \citenamefont {Maiorano}, \citenamefont {Gigli}, \citenamefont {Gerace}, \citenamefont {Sanvitto},\ and\ \citenamefont {Ballarini}}]{Maggiolini2023}%
  \BibitemOpen
  \bibfield  {author} {\bibinfo {author} {\bibfnamefont {E.}~\bibnamefont {Maggiolini}}, \bibinfo {author} {\bibfnamefont {L.}~\bibnamefont {Polimeno}}, \bibinfo {author} {\bibfnamefont {F.}~\bibnamefont {Todisco}}, \bibinfo {author} {\bibfnamefont {A.}~\bibnamefont {{Di Renzo}}}, \bibinfo {author} {\bibfnamefont {B.}~\bibnamefont {Han}}, \bibinfo {author} {\bibfnamefont {M.}~\bibnamefont {{De Giorgi}}}, \bibinfo {author} {\bibfnamefont {V.}~\bibnamefont {Ardizzone}}, \bibinfo {author} {\bibfnamefont {C.}~\bibnamefont {Schneider}}, \bibinfo {author} {\bibfnamefont {R.}~\bibnamefont {Mastria}}, \bibinfo {author} {\bibfnamefont {A.}~\bibnamefont {Cannavale}}, \bibinfo {author} {\bibfnamefont {M.}~\bibnamefont {Pugliese}}, \bibinfo {author} {\bibfnamefont {L.}~\bibnamefont {{De Marco}}}, \bibinfo {author} {\bibfnamefont {A.}~\bibnamefont {Rizzo}}, \bibinfo {author} {\bibfnamefont {V.}~\bibnamefont {Maiorano}}, \bibinfo {author} {\bibfnamefont {G.}~\bibnamefont {Gigli}}, \bibinfo {author} {\bibfnamefont
  {D.}~\bibnamefont {Gerace}}, \bibinfo {author} {\bibfnamefont {D.}~\bibnamefont {Sanvitto}},\ and\ \bibinfo {author} {\bibfnamefont {D.}~\bibnamefont {Ballarini}},\ }\bibfield  {title} {\bibinfo {title} {{Strongly enhanced light–matter coupling of monolayer WS2 from a bound state in the continuum}},\ }\href {https://doi.org/10.1038/s41563-023-01562-9} {\bibfield  {journal} {\bibinfo  {journal} {Nature Materials}\ }\textbf {\bibinfo {volume} {22}},\ \bibinfo {pages} {964} (\bibinfo {year} {2023})}\BibitemShut {NoStop}%
\bibitem [{\citenamefont {Wu}\ \emph {et~al.}(2023)\citenamefont {Wu}, \citenamefont {Song}, \citenamefont {Zhang}, \citenamefont {Du}, \citenamefont {Wang}, \citenamefont {Zhu}, \citenamefont {Liang}, \citenamefont {Zhang}, \citenamefont {Xiong},\ and\ \citenamefont {Liu}}]{wu2023roomtemperature}%
  \BibitemOpen
  \bibfield  {author} {\bibinfo {author} {\bibfnamefont {X.}~\bibnamefont {Wu}}, \bibinfo {author} {\bibfnamefont {J.}~\bibnamefont {Song}}, \bibinfo {author} {\bibfnamefont {S.}~\bibnamefont {Zhang}}, \bibinfo {author} {\bibfnamefont {W.}~\bibnamefont {Du}}, \bibinfo {author} {\bibfnamefont {Y.}~\bibnamefont {Wang}}, \bibinfo {author} {\bibfnamefont {Z.}~\bibnamefont {Zhu}}, \bibinfo {author} {\bibfnamefont {Y.}~\bibnamefont {Liang}}, \bibinfo {author} {\bibfnamefont {Q.}~\bibnamefont {Zhang}}, \bibinfo {author} {\bibfnamefont {Q.}~\bibnamefont {Xiong}},\ and\ \bibinfo {author} {\bibfnamefont {X.}~\bibnamefont {Liu}},\ }\href@noop {} {\bibinfo {title} {Room-temperature bound states in the continuum polariton condensation}} (\bibinfo {year} {2023}),\ \Eprint {https://arxiv.org/abs/2303.09923} {arXiv:2303.09923 [cond-mat.mes-hall]} \BibitemShut {NoStop}%
\bibitem [{\citenamefont {Mermet-Lyaudoz}\ \emph {et~al.}(2023)\citenamefont {Mermet-Lyaudoz}, \citenamefont {Symonds}, \citenamefont {Berry}, \citenamefont {Drouard}, \citenamefont {Chevalier}, \citenamefont {Trippé-Allard}, \citenamefont {Deleporte}, \citenamefont {Bellessa}, \citenamefont {Seassal},\ and\ \citenamefont {Nguyen}}]{Mermet2023}%
  \BibitemOpen
  \bibfield  {author} {\bibinfo {author} {\bibfnamefont {R.}~\bibnamefont {Mermet-Lyaudoz}}, \bibinfo {author} {\bibfnamefont {C.}~\bibnamefont {Symonds}}, \bibinfo {author} {\bibfnamefont {F.}~\bibnamefont {Berry}}, \bibinfo {author} {\bibfnamefont {E.}~\bibnamefont {Drouard}}, \bibinfo {author} {\bibfnamefont {C.}~\bibnamefont {Chevalier}}, \bibinfo {author} {\bibfnamefont {G.}~\bibnamefont {Trippé-Allard}}, \bibinfo {author} {\bibfnamefont {E.}~\bibnamefont {Deleporte}}, \bibinfo {author} {\bibfnamefont {J.}~\bibnamefont {Bellessa}}, \bibinfo {author} {\bibfnamefont {C.}~\bibnamefont {Seassal}},\ and\ \bibinfo {author} {\bibfnamefont {H.~S.}\ \bibnamefont {Nguyen}},\ }\bibfield  {title} {\bibinfo {title} {Taming friedrich–wintgen interference in a resonant metasurface: Vortex laser emitting at an on-demand tilted angle},\ }\href {https://doi.org/10.1021/acs.nanolett.2c04936} {\bibfield  {journal} {\bibinfo  {journal} {Nano Letters}\ }\textbf {\bibinfo {volume} {23}},\ \bibinfo {pages} {4152} (\bibinfo
  {year} {2023})}\BibitemShut {NoStop}%
\bibitem [{\citenamefont {Liu}\ \emph {et~al.}(2015)\citenamefont {Liu}, \citenamefont {Kleimann}, \citenamefont {Laffite}, \citenamefont {Jamois},\ and\ \citenamefont {Orobtchouk}}]{Liu2015}%
  \BibitemOpen
  \bibfield  {author} {\bibinfo {author} {\bibfnamefont {J.}~\bibnamefont {Liu}}, \bibinfo {author} {\bibfnamefont {P.}~\bibnamefont {Kleimann}}, \bibinfo {author} {\bibfnamefont {G.}~\bibnamefont {Laffite}}, \bibinfo {author} {\bibfnamefont {C.}~\bibnamefont {Jamois}},\ and\ \bibinfo {author} {\bibfnamefont {R.}~\bibnamefont {Orobtchouk}},\ }\bibfield  {title} {\bibinfo {title} {Formation of 300 nm period pore arrays by laser interference lithography and electrochemical etching},\ }\href {https://doi.org/10.1063/1.4907620} {\bibfield  {journal} {\bibinfo  {journal} {Applied Physics Letters}\ }\textbf {\bibinfo {volume} {106}},\ \bibinfo {pages} {053107} (\bibinfo {year} {2015})},\ \Eprint {https://arxiv.org/abs/https://pubs.aip.org/aip/apl/article-pdf/doi/10.1063/1.4907620/14464895/053107\_1\_online.pdf} {https://pubs.aip.org/aip/apl/article-pdf/doi/10.1063/1.4907620/14464895/053107\_1\_online.pdf} \BibitemShut {NoStop}%
\bibitem [{\citenamefont {Friedrich}\ and\ \citenamefont {Wintgen}(1985)}]{FW85}%
  \BibitemOpen
  \bibfield  {author} {\bibinfo {author} {\bibfnamefont {H.}~\bibnamefont {Friedrich}}\ and\ \bibinfo {author} {\bibfnamefont {D.}~\bibnamefont {Wintgen}},\ }\bibfield  {title} {\bibinfo {title} {Interfering resonances and bound states in the continuum},\ }\href {https://doi.org/10.1103/PhysRevA.32.3231} {\bibfield  {journal} {\bibinfo  {journal} {Phys. Rev. A}\ }\textbf {\bibinfo {volume} {32}},\ \bibinfo {pages} {3231} (\bibinfo {year} {1985})}\BibitemShut {NoStop}%
\bibitem [{\citenamefont {Hsu}\ \emph {et~al.}(2016)\citenamefont {Hsu}, \citenamefont {Zhen}, \citenamefont {Stone}, \citenamefont {Joannopoulos},\ and\ \citenamefont {Solja{\v{c}}i{\'{c}}}}]{Hsu2016}%
  \BibitemOpen
  \bibfield  {author} {\bibinfo {author} {\bibfnamefont {C.~W.}\ \bibnamefont {Hsu}}, \bibinfo {author} {\bibfnamefont {B.}~\bibnamefont {Zhen}}, \bibinfo {author} {\bibfnamefont {A.~D.}\ \bibnamefont {Stone}}, \bibinfo {author} {\bibfnamefont {J.~D.}\ \bibnamefont {Joannopoulos}},\ and\ \bibinfo {author} {\bibfnamefont {M.}~\bibnamefont {Solja{\v{c}}i{\'{c}}}},\ }\bibfield  {title} {\bibinfo {title} {{Bound states in the continuum}},\ }\href {https://doi.org/10.1038/natrevmats.2016.48} {\bibfield  {journal} {\bibinfo  {journal} {Nature Reviews Materials}\ }\textbf {\bibinfo {volume} {1}},\ \bibinfo {pages} {16048} (\bibinfo {year} {2016})}\BibitemShut {NoStop}%
\bibitem [{\citenamefont {Gerace}\ and\ \citenamefont {Andreani}(2007)}]{Dario2007}%
  \BibitemOpen
  \bibfield  {author} {\bibinfo {author} {\bibfnamefont {D.}~\bibnamefont {Gerace}}\ and\ \bibinfo {author} {\bibfnamefont {L.~C.}\ \bibnamefont {Andreani}},\ }\bibfield  {title} {\bibinfo {title} {Quantum theory of exciton-photon coupling in photonic crystal slabs with embedded quantum wells},\ }\href {https://doi.org/10.1103/PhysRevB.75.235325} {\bibfield  {journal} {\bibinfo  {journal} {Phys. Rev. B}\ }\textbf {\bibinfo {volume} {75}},\ \bibinfo {pages} {235325} (\bibinfo {year} {2007})}\BibitemShut {NoStop}%
\bibitem [{\citenamefont {Zanotti}\ \emph {et~al.}(2022)\citenamefont {Zanotti}, \citenamefont {Nguyen}, \citenamefont {Minkov}, \citenamefont {Andreani},\ and\ \citenamefont {Gerace}}]{Zanotti2022}%
  \BibitemOpen
  \bibfield  {author} {\bibinfo {author} {\bibfnamefont {S.}~\bibnamefont {Zanotti}}, \bibinfo {author} {\bibfnamefont {H.~S.}\ \bibnamefont {Nguyen}}, \bibinfo {author} {\bibfnamefont {M.}~\bibnamefont {Minkov}}, \bibinfo {author} {\bibfnamefont {L.~C.}\ \bibnamefont {Andreani}},\ and\ \bibinfo {author} {\bibfnamefont {D.}~\bibnamefont {Gerace}},\ }\bibfield  {title} {\bibinfo {title} {Theory of photonic crystal polaritons in periodically patterned multilayer waveguides},\ }\href {https://doi.org/10.1103/PhysRevB.106.115424} {\bibfield  {journal} {\bibinfo  {journal} {Phys. Rev. B}\ }\textbf {\bibinfo {volume} {106}},\ \bibinfo {pages} {115424} (\bibinfo {year} {2022})}\BibitemShut {NoStop}%
\end{thebibliography}%

\end{document}